\documentclass[useAMS,usenatbib]{mn2e}

\usepackage[dvips]{graphicx}
\usepackage{graphicx,subfigure}
\usepackage{rotating}
\usepackage{amssymb,amsmath}
\usepackage{url}

\title[Binaries of MBHs in rotating clusters: 
       Dynamics, GWs, detection and eccentricity]
      {Binaries of massive black holes in rotating clusters:\\
       Dynamics, gravitational waves, detection
       and the role of eccentricity}

\author[P. Amaro-Seoane, C. Eichhorn, E. K. Porter and R. Spurzem]
{
P. Amaro-Seoane$^{1,2}$\thanks{E-mail: Pau.Amaro-Seoane@aei.mpg.de}, 
C. Eichhorn$^{3}$\thanks{E-mail: Eichhorn@irs.uni-stuttgart.de}, 
Edward K. Porter$^{4}$\thanks{E-mail: porter@apc.univ-paris7.fr},
R. Spurzem$^{5,6,7}$\thanks{E-mail: Spurzem@ari.uni-heidelberg.de}\\
$^{1}$Max-Planck Institut f{\"u}r Gravitationsphysik (Albert Einstein-Institut), 
Am M{\"u}hlenberg 1, D-14476 Potsdam, Germany\\ 
$^{2}$Institut de Ci{\`e}ncies de l'Espai, IEEC/CSIC, Campus UAB, Torre C-5, parells, 
$2^{\rm na}$ planta, ES-08193, Bellaterra, Barcelona, Spain\\
$^{3}$Institut f{\"u}r Raumfahrtsysteme, Universit{\"a}t Stuttgart, Pfaffenwaldring 31, 70550 Stuttgart, Germany\\
$^{4}$Laboratoire APC, UMR 7164, Universit\'e Paris 7 Denis Diderot, 10, rue Alice Domon et L{\'e}onie Duquet 75205 Paris Cedex 13, France\\
$^{5}$National Astronomical Observatories of China, Chinese Academy of
Sciences, 20A Datun Lu, Chaoyang District, 100012, Beijing, China\\
$^{6}$Kavli Institute for Astronomy and Astrophysics, Peking University, China\\
$^{7}$Astronomisches Rechen-Institut, M{\"o}nchhofstra{\ss}e 12-14, 69120, 
Zentrum f\"ur Astronomie, Universit\"at Heidelberg, Germany
}

\begin{document}

\date{\today}

\pagerange{\pageref{firstpage}--\pageref{lastpage}} \pubyear{2009}

\maketitle

\label{firstpage}

\begin{abstract}

The dynamical evolution of binaries of intermediate-massive black holes (IMBHs,
massive black holes with a mass ranging between $10^2$ and $10^4\,M_{\odot}$)
in stellar clusters has recently received an increasing amount of attention.
This is at least partially due to the fact that if the binary is hard enough to
evolve to the phase at which it will start emitting gravitational waves
(GWs) efficiently, there is a good probability that it will be detectable by future
space-borne detectors like LISA.
We study this evolution in the presence of rotation in the cluster
by carrying out a series of simulations of an
equal-mass binary of IMBHs embedded in a stellar distribution with different
rotational parameters. 
The survey indicates that eccentricities and inclinations 
are primarily determined by the initial conditions
of the IMBHs and the influence of dynamical friction, even though they are finally
perturbed by the scattering of field stars. In particular, the eccentricity is
strongly connected to the initial IMBHs velocities, and values of $\sim 0.7$ up
to $0.9$ are reached for low initial velocities, while almost circular orbits
result if the initial velocities are increased. Evidence suggests a dependency
of the eccentricity on the rotation parameter. We found only weak changes in the
inclination, with slight variations of the orientation of the
angular momentum vector of the binary. Counter-rotation simulations
yield remarkably different results in eccentricity.
A Monte Carlo study indicates that these sources will be detectable by a detector such as LISA with median signal to noise ratios of
between 10 and 20 over a three year period, although some events had signal to noise ratios of 300 or
greater.  Furthermore, one should also be able to estimate the chirp-mass with
median fractional errors of $10^{-4}$, reduced mass on the order of $10^{-3}$
and luminosity distance on the order of $10^{-1}$. Finally, these sources will have a median angular resolution in the
LISA detector of about 3 square degrees, putting events firmly in the field of
view of future electromagnetic detectors such as LSST.

\end{abstract}


\section{Introduction}

Even though their existence are not as well-established as the existence of
stellar-mass or supermassive black holes (SMBHs), it is plausible that
intermediate-mass black holes (IMBHs; masses $M\sim 10^{2-4}~M_\odot$) exist in
the centre of stellar clusters \citep[see][and references
therein]{MillerColbert04}. The coalescence of a binary of two massive black
holes in that mass range is a powerful source of gravity waves which will be
detectable by space-born observatories such as the Laser Interferometer Space Antenna (LISA\footnote{\url{http://lisa.nasa.gov/}, \url{http://sci.esa.int/science-e/www/area/index.cfm?fareaid=27}}) \citep{ASF06,Amaro-SeoaneEtAl09a}.  

There are two possible ways to theoretically explain the presence of a binary
of IMBHs in a globular cluster. (1) The ``single cluster channel'': \cite{GFR06} address this possibility in the
scenario of a runaway growth of two stars in a young cluster via physical
collisions in the innermost part of the stellar system, where the heaviest
stars have sunk through mass segregation
\citep{PortegiesZwartMcMillan00,GurkanEtAl04,PortegiesZwartEtAl04,
FreitagEtAl06}. By adding a fraction of primordial binaries to the cluster,
\cite{GFR06} found not one but two independent very massive stars growing in
the centre. Eventually, and due to post-Newtonian inestabilities, they could
collapse and form a massive black hole in the relevant mass range.
\cite{Amaro-SeoaneEtAl09a} followed the evolution of such a binary with direct-summation
$N-$body simulations and estimated that, in some cases, one can expect a
residual eccentricity as high as 0.3 for the binary when it enters the LISA
band. (2) The ``double cluster channel'': \cite{ASF06} discussed an alternative way to form binaries of
IMBHs in stellar clusters. It has been observed that in star forming regions
such as the Antenn{\ae} or Stephan's Quintet, hundreds of young massive star
clusters are clustered into larger complexes of a few 100 pc across
\citep{WhitmoreEtAl99,ZhangFall99,GallagheretAl01}. These clusters contain
typically some $\sim 10^5$ stars within a few parsecs and it is most likely
that some of them are bound \citep{DMG02}. Also, most of the clusters in
binaries are coeval and younger than 300\,Myr, which means that they merge
early. \cite{ASF06} followed the evolution of two IMBHs initially located at the
centre of two such clusters which collide, and studied the orbital decay
to the centre.  

The evolution of the orbital parameters of the binary of massive BHs is
determined by the stellar dynamics and the emission of GWs, which is negligible
at long distances but becomes more and more dominant as the semi-major axis of
the binary shrinks.  In order to understand the distribution of parameters we
can expect for these systems when they enter the window of detection of LISA,
it is important to analyse the dynamical story of the binary in detail. The
evolution starting at distances at which dynamical friction is important down
to the phase of strong emission of GWs --and their detection and
characterisation-- is a complex and long process which requires different
techniques to address it. 

One can distinguish roughly three different regimes in the process: (i) at the
beginning, the massive BHs are at distances in which GWs, though always
present, are totally negligible and the evolution is dominated purely by the
dynamics. At this stage, dynamical friction will sink the two massive BHs down
to the centre. The perturbers, the single massive BHs in our case, are moving
through a sea of small stars (as compared to their masses).  The velocity
vector of the stars are rotated after deflecting with the IMBHs.  The projected
component in the direction of the deflection is shorter. Hence, the massive
object, the IMBH, is cumulating just after it a high-density stellar region.
The perturber will feel a drag from that region from the conservation of $J$ in
the direction of its velocity vector. The direction does not change to
first-order, but the amplitude decreases A drag force starts to act on to the
perturber, so that it slows downs as it sinks down to the centre of the stellar
system. This force happens to be proportional the square of the mass of the
perturber so, the bigger the mass of the perturber, the bigger the dynamical
effects, in spite of the bigger inertia. (ii) As they approach closer and
closer, the two massive BHs form a bound state, a binary system. At this stage,
the binary interacts strongly with stars coming from the surrounding stellar
system. Since the stars have a much smaller mass, the outcome of the
interaction is that a star is slingshot into the stellar system with a higher
kinetic energy, gained from the removal of energy and angular momentum of the
binary; therefore, the semi-major axis of the binary, $a$, shrinks a bit more.
These interactions mostly tend to increase the eccentricity of the binary (iii)
The process is repeated again and again, provided the reservoir of stars is not
empty -- in which case the loss-cone would be depleted --, until the separation
between the members of the binary is small enough that the emission of GWs is
strong enough as to take over the dynamics as the main factor of evolution of
the orbital parameters. The binary is practically isolated from the stellar
system. Thereafter, the binary begins to circularise.  Obviously, the
transition between these phases in the evolution is not well-separated and the
whole evolution requires numerical tools to investigate it.
\cite{ASF06,Amaro-SeoaneEtAl09a} addressed some of these questions in the
context of globular clusters and IMBHs.  They proved, with $N-$body models,
that slingshot ejections of stars increase the eccentricity of the IMBH binary
to $\sim 0.8$ and beyond and that later the emission of GWs then circularises
the orbit to rather low, yet detectable values of eccentricity. 

But Nature is more complex than that.  A key effect that will determine the
global evolution of both the cluster and of the massive binary of IMBHs is the
rotation of the stellar system.  Rotation has been identified in clusters for a
long time. Deviations from spherical symmetry were discovered in some globular
clusters \citep{Shapley30} as early as the beginning of the last century. This
flattening is a fingerprint for rotation and the measurements of ellipticity
were later extended to galactic and extragalactic globular clusters \citep[see
e.g.][]{WhiteShawl87,StanevaEtAl96}. One can also detect rotation by measuring
radial velocities of individual stars in the globular clusters
\citep{MeylanMayor86,GebhardtEtAl00,ReijnsEtAl06}.  

\cite{ASF06} first addressed this problem focusing on the detection of GWs in
the context of a binary of IMBH formed as the result of a collision of two
clusters; \cite{Amaro-SeoaneEtAl09a} looked at the same problem from the
perspective of a born-in binary, extended the study to multi-mass clusters and
non-equal mass binaries and described the global dynamical evolution with the
implication of the larger eccentricities they found in a general context of
detection.

The role of the eccentricity of the binary is decisive in the study.  Direct
summation $N$-body models, including post-Newtonian corrections, show that the
stellar dynamical history before the relativistic regime can significantly
affect the final evolution and leads to different merger times \citep{BerentzenEtAl09}.
In particular, it turns out that massive binaries may enter the relativistic
phase with high eccentricities, and signatures of the eccentricities are kept in
the harmonics of the gravitational waveforms until the moment of coalescence
\citep{ASF06,Amaro-SeoaneEtAl09a,BerentzenEtAl09}.  The evolution of the eccentricity
has been previously discussed in a number of articles
\citep{mak1993,hem2000,mil2001,hem2002,BerczikEtAl05,ASF06,bms2006,Amaro-SeoaneEtAl09a,BerentzenEtAl09}. 

As derived by \citet{pet1964}, the orbit-averaged timescale of coalescence due
to the emission of gravitational radiation is given by

\begin{equation}
t_{\rm GW}=\frac{5}{64}
\frac{c^{5}a_{\rm GW}^{4}}{G^{3}M_{1}M_{2}(M_{1}+M_{2})F(e)}
\end{equation}

\noindent
where $M_{1}$, $M_{2}$ denote the black hole masses, $a_{\rm GW}$
the characteristic separation for gravitational wave emission, $G$
the gravitational constant, $c$ the speed of light and

\begin{equation}
F(e)=(1-e^{2})^{-7/2}\left(1+\frac{73}{24}e^{2}+\frac{37}{96}e^{4}\right)
\end{equation}

\noindent
is a function with dependence on the eccentricity $e$. Thus the coalescence time
can shrink by several orders of magnitude if the eccentricity is high enough,
resulting in a stronger burst of gravitational radiation and different
characteristic waveforms.  Furthermore, the behavior of the the inclination of
the orbital plane is potentially interesting in predicting processes related to
angular momentum exchange between IMBHs and field stars.

A natural continuation of the analysis carried out until now is to add another
physical factor to the problem, the rotation of the system, since it can have a
very important impact in the global dynamics of the cluster. It can also
particularly effect the evolution of the eccentricity of the binary and, thus, the detection and characterisation of the GW observation.

From a standpoint of the data analysis of such GWs for LISA, because of their low mass
--as compared to SMBH binaries--, IMBH binaries should be visible at moderate to
high frequencies in the LISA detector.  As most of these binaries coalesce
outside the LISA band, they will be observable in the detector throughout the
lifetime of the mission.  This should allow us to confidently detect and
estimate the parameters for such systems.  If such sources exist in the LISA
data stream, it will also allow us, assuming a strong enough signal, to provide
accurate distance measurements in the local universe.

The stucture of this paper is as follows: We start by giving a description of
the numerical method used for the simulations in section\,(\ref{sec.method});
later, in section\,(\ref{sec.initial}) we give a short overview of the initial
conditions we use for the numerical study; in section\,(\ref{sec.dynamics1}) we
provide a detailed analysis of the dynamics of the system, i.e. evolution of
the binding energy, eccentricity and inclination of the IMBH binary; in
section\,(\ref{sec.dynamics2}) we study some cases in which the binary is
initially set up on a counter-rotating orbit in relation to the stellar system
in which it is embedded and study the associated Brownian motion; in
section\,(\ref{sec.GW}) we dicuss about the implications for lower-frequency
Astrophysics and the detectability of such systems. In the last
section\,(\ref{sec.conclusions}) we summarise the results and give the
conclusions of the study.

\section{Numerical method}
\label{sec.method}

The simulations in this work have been performed using NBODY6++, a parallelised
version of Aarseth's NBODY6 \citep{spu1999}. The code includes a
Hermite integration scheme, KS-regularisation \citep{kus1965} and
the Ahmad-Cohen neighbour scheme \citep{ahm1973}. No softening has
been introduced; this circumstance allows an accurate treatment of
the effects due to super-elastic scattering events, which play a
crucial part in black hole binary evolution and require a precise
calculation of the trajectories throughout the interaction.

Additionally, in order to improve the exactness of computation of the motion of
particles in the environment of the black holes, a modification in the
determination of the neighbour radius in the Ahmad-Cohen scheme was
implemented. In principle, the Ahmad-Cohen scheme divides the force on a
particle into a regular and an irregular component, assigning both forces
different time steps; the irregular component is computed over the nearest
particles populating an area called the neighbour sphere. Normally, the
neighbour sphere, which is in its extension defined by the neighbour radius, is
characterised as containing a defined number of stars $n_{n}$, which is
typically set to $n_{n}=50$ in simulations dealing with particle numbers of the
order presented here. However, when considering a scenario consisting of two
heavy particles of the same mass, embedded in a stellar component of equal-mass
particles, the neighbour radius is enlarged by a factor

\begin{equation}
\gamma=\beta\left[\left(\frac{1}{2}\left(\frac{m_{i}}{m_{j}} +
\frac{m_{j}}{m_{i}}\right)\right)^{\lambda}-1\right]+1
\end{equation}

\noindent
if during the declaration of the neighbour particles of a particle $j$ and the
neighbour candidates $i$ a mass difference $m_{i}/m_{j}\neq 1$ occurs
\citep{hem2000}. The enlargement factor $\gamma$ is symmetric in the masses
$m_{i}$ and $m_{j}$. As a result, massive particles (black holes) receive a
larger neighbour radius, and a massive particle is also more likely declared as
neighbour of a stellar particle. This method accommodates the influence of a
black hole on its surroundings and lessens the underestimation on that stellar
component which possessed a black hole nearby, but outside the neighbour sphere
in the scheme without the enlargement factor. The parameters $\beta$ and
$\lambda$ have been set $\beta=0.03$ and $\lambda=1$, yielding $\gamma=10.57$,
consistent with the simulations presented by \citet{hem2002}.

\section{Initial conditions}
\label{sec.initial}

In all our simulations, the conversion factors are as follows: A unit of length
is equivalent to ${\cal U}|^{\rm NB}_{\rm L} = 1.1$ pc, a unit of mass to
${\cal U}|^{\rm NB}_{\rm M} = 43921.6 M_{\odot}$, a unit of velocity to ${\cal
U}|^{\rm NB}_{\rm V} =  13.233$ km/s and a unit of time to  ${\cal U}|^{\rm
NB}_{\rm M} = 8.14\cdot10^{-2}$ Myrs. The set of simulations was carried out
for a total particle number $N=64 000$, including two massive black holes
$M_{1}=M_{2}=0.01$ embedded in a dense stellar system of 63998 equal-mass
particles $m_{\star}=1.5625\cdot 10^{-5}\,{\cal U}|^{\rm NB}_{\rm M}$. The initial
stellar distribution satisfies a rotating King model,

\begin{equation}
f(E,L_{z})= \left\{
\begin{array}{cc}
\mathrm{const}\times
\left(e^{-\beta(E-\Phi_{t})}-1\right)e^{\beta\Omega_{c} L_{z}}
&  E<\Phi_{t} \\
0 &  E\geq\Phi_{t} \\
\end{array}
\right.
\end{equation}

\noindent
where $\beta=1/\sigma^{2}$ represents the inverse one-dimensional velocity
dispersion, $E$ and $L_{z}$ the energy and the z-component of the angular
momentum of a star per unit mass, and $\Phi_{t}$ the tidal potential of the
model. The King parameters $W_{0}$ and $\omega_{0}$ are defined by
$W_{0}=-\beta(\Phi_{0}-\Phi_{t})$ and $\omega_{0}=\sqrt{9/4\pi G\rho_{c}}\cdot
\Omega_{c}$ with the parameters $\Phi_{0}$ representing the central potential,
$\rho_{c}$ a mass density and $\Omega_{c}$ approximately the angular velocity
in the centre \citep{ein1999,lag1996,lon1996}.

\begin{table}
\begin{tabular}{c|c|c|c|c|c|c|c}
\hline Set & Model & $W_{0}$ & $\omega_{0}$ & $v_{0}$ & $r_{c}$ & $v_{c}$ & $T_{\rm rot}/T_{\rm kin}$\\
\hline 
  & ${\cal A}$ & 3 & 0.0 & $\sqrt{2}v_{c}$ & 0.396 & 0.552 & 0\\
1 & ${\cal B}$ & 3 & 0.0 & $v_{c}$         & 0.396 & 0.552 & 0\\
  & ${\cal C}$ & 3 & 0.0 & $0.136v_{c}$    & 0.396 & 0.552 & 0\\
\hline
  & ${\cal D}$ & 3 & 0.3 & $\sqrt{2}v_{c}$ & 0.400 & 0.558 & 0.01\\
2 & ${\cal E}$ & 3 & 0.3 & $v_{c}$         & 0.400 & 0.558 & 0.01\\
  & ${\cal F}$ & 3 & 0.3 & $0.136v_{c}$    & 0.400 & 0.558 & 0.01\\
\hline
  & ${\cal G}$ & 3 & 0.6 & $\sqrt{2}v_{c}$ & 0.415 & 0.555 & 0.0343\\
3 & ${\cal H}$ & 3 & 0.6 & $v_{c}$         & 0.415 & 0.555 & 0.0343\\
  & ${\cal I}$ & 3 & 0.6 & $0.136v_{c}$    & 0.415 & 0.555 & 0.0343\\
\hline
  & ${\cal J}$ & 6 & 0.0 & $\sqrt{2}v_{c}$ & 0.239 & 0.563 & 0\\
4 & ${\cal K}$ & 6 & 0.0 & $v_{c}$         & 0.239 & 0.563 & 0\\
  & ${\cal L}$ & 6 & 0.0 & $0.136v_{c}$    & 0.239 & 0.563 & 0\\
\hline
  & ${\cal M}$ & 6 & 0.3 & $\sqrt{2}v_{c}$ & 0.256 & 0.553 & 7\\
5 & ${\cal N}$ & 6 & 0.3 & $v_{c}$         & 0.256 & 0.553 & 7\\
  & ${\cal O}$ & 6 & 0.3 & $0.136v_{c}$    & 0.256 & 0.553 & 7\\
\hline
6 & ${\cal P}$ & 6 & 0.6 & $v_{c}$         & 0.320 & 0.550 & 19.81\\
  & ${\cal Q}$ & 6 & 0.6 & $0.136v_{c}$    & 0.320 & 0.550 & 19.81\\
\hline
\end{tabular}
\caption{
Overview over the initial conditions of all simulations. All values are
expressed in N-body units. $W_0$ is the King parameter, $\omega_{0}$ the
rotation parameter, $v_{0}$ the initial velocity of the IMBHs, $r_{c}$ the core
radius, $v_{c}$ the central velocity and $T_{\rm rot}/T_{\rm kin}$ is the ratio
of rotational energy and total kinetic energy at time $T = 0$.  We use $\gamma
= \gamma_3$, $\beta = 0.03$, $\lambda = 1.0$. In all simulations ${\eta}_{\rm
reg}=0.02$, whilst ${\eta}_{\rm irreg}=0.02$ for the cases ${\cal A} - {\cal
I}$ and 0.01 for the cases ${\cal J} - {\cal Q}$
}
\label{initialcond}
\end{table}
A symmetric set-up underlies all simulations, with the rotation axis of the
King model in the z-direction and the two black holes located in the xy-plane
on the opposite side of the cluster on the core radius $r_{c}$ of the model.
Thereby, the usual definition of the core radius in N-body simulations,

\begin{equation}\label{10}
r_{c}=\left(\sum_{j=1}^{N^{*}}\rho_{j}^{2}|\textbf{r}_{j}-\textbf{r}_{d}|^{2}
/\sum_{j=1}^{N^{*}}\rho_{j}^{2}\right)^{1/2} \qquad N^{*}\geq N/5,
\end{equation}
is used \citep{aar2003}, where
$\textbf{r}_{d}=\sum_{j=1}^{N}\rho_{j}\textbf{r}_{j}/\sum_{j=1}^{N}\rho_{j}$
is the density centre of the system consisting of particles of the
mass $m$ with the coordinates $\textbf{r}_{j}$, and
$\rho_{j}=3(k-1)m/4\pi r_{j(k)}^{3}$ the local density in an area
around each particle $j$ \citep{cas1985}. The quantity $r_{j(k)}$
is to be interpreted as the radius of the sphere over which the
local density is evaluated, characterised in size by the number of
stars $k$ populating this volume. Applying this formalism, $k=6$
is the optimal choice. The summation in equation \ref{10} is restricted to
the innermost $N/5$ particles, which saves computation time while
showing coevally adequate agreement with an exact calculations.
The initial velocities are adjusted in the xy-plane tangential to
a circle around the centre, adopting the values $v_{0}=0.136v_{c},
v_{c}$ and $\sqrt{2}v_{c}$ in terms of the circular velocity
$v_{c}$. Scenarios with the black holes moving initially with, as
well as contrary to, the rotation of the stellar system have been
investigated.

The evolution of the black hole binary was followed using King parameters
$W_{0}=3$; $6$ and $\omega_{0}=0.0; 0.3; 0.6$.  We have chosen these parameters
because they are representative for the problem we want to address in this
work, from the absence of rotation to a rather high value. Typical values for
$\omega_{0}$ in globular clusters range between 0.1 and 0.5.  For instance,
$\omega Cen$ has a value of 0.5, N2808 of 0.3, 47Tuc of 0.15, N5286 of 0.5 etc
\citep{FiestasEtAl06}. 

The accuracy was tuned in such a way that relative energy errors measured over
one NBODY time were $<10^{-3}$ concerning $W_{0}=3$ and $<10^{-4}$ concerning
$W_{0}=6$ simulations respectively. The complete survey of investigations is
shown in Tab.\ref{initialcond}. 

\section{Dynamics of the system}
\label{sec.dynamics1}

\subsection{Evolution of the binding energy}
\label{sec.EvolBindEnergy}

\begin{figure*}
\includegraphics[width=6.0cm,angle=0]{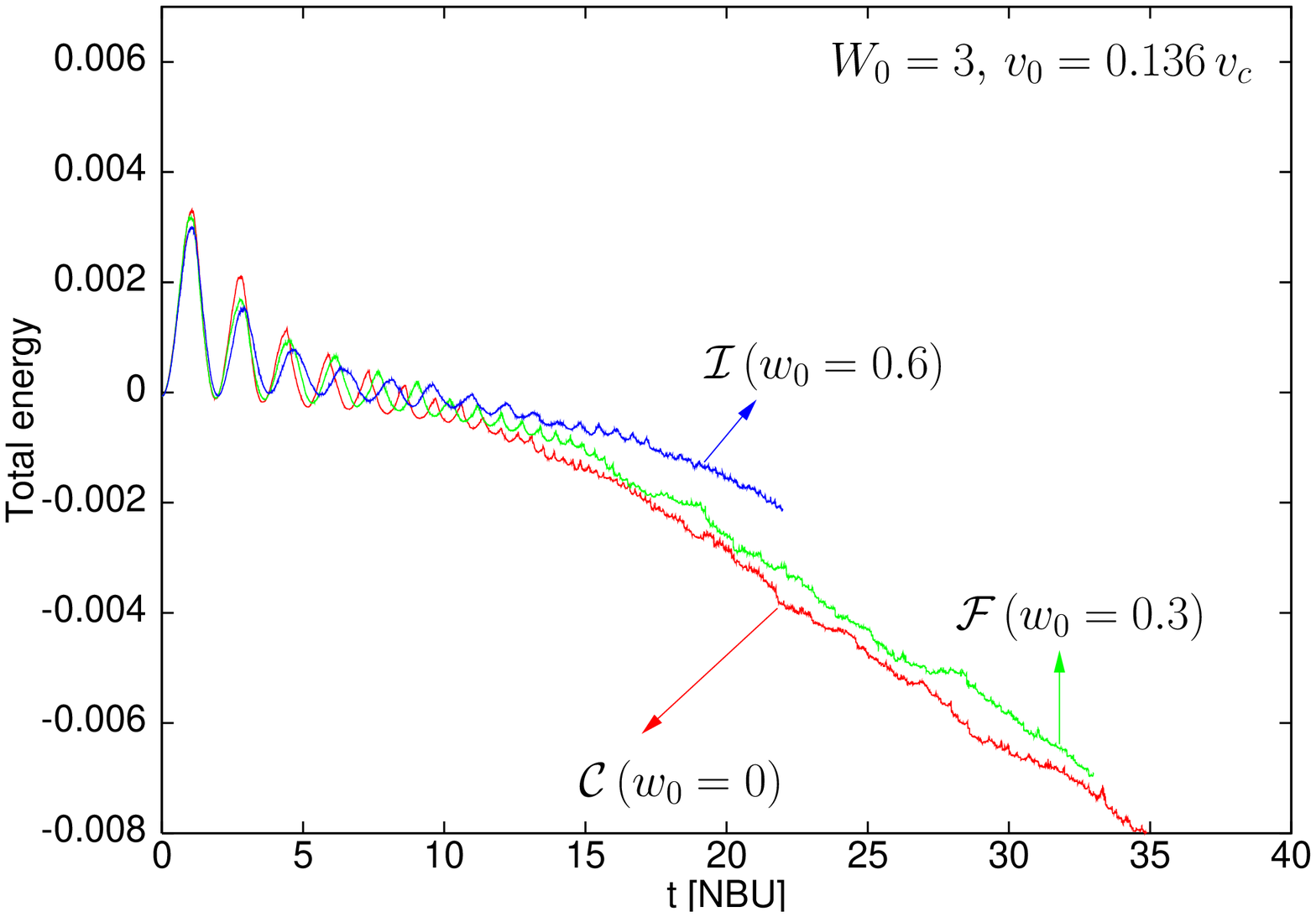}
\includegraphics[width=6.0cm,angle=0]{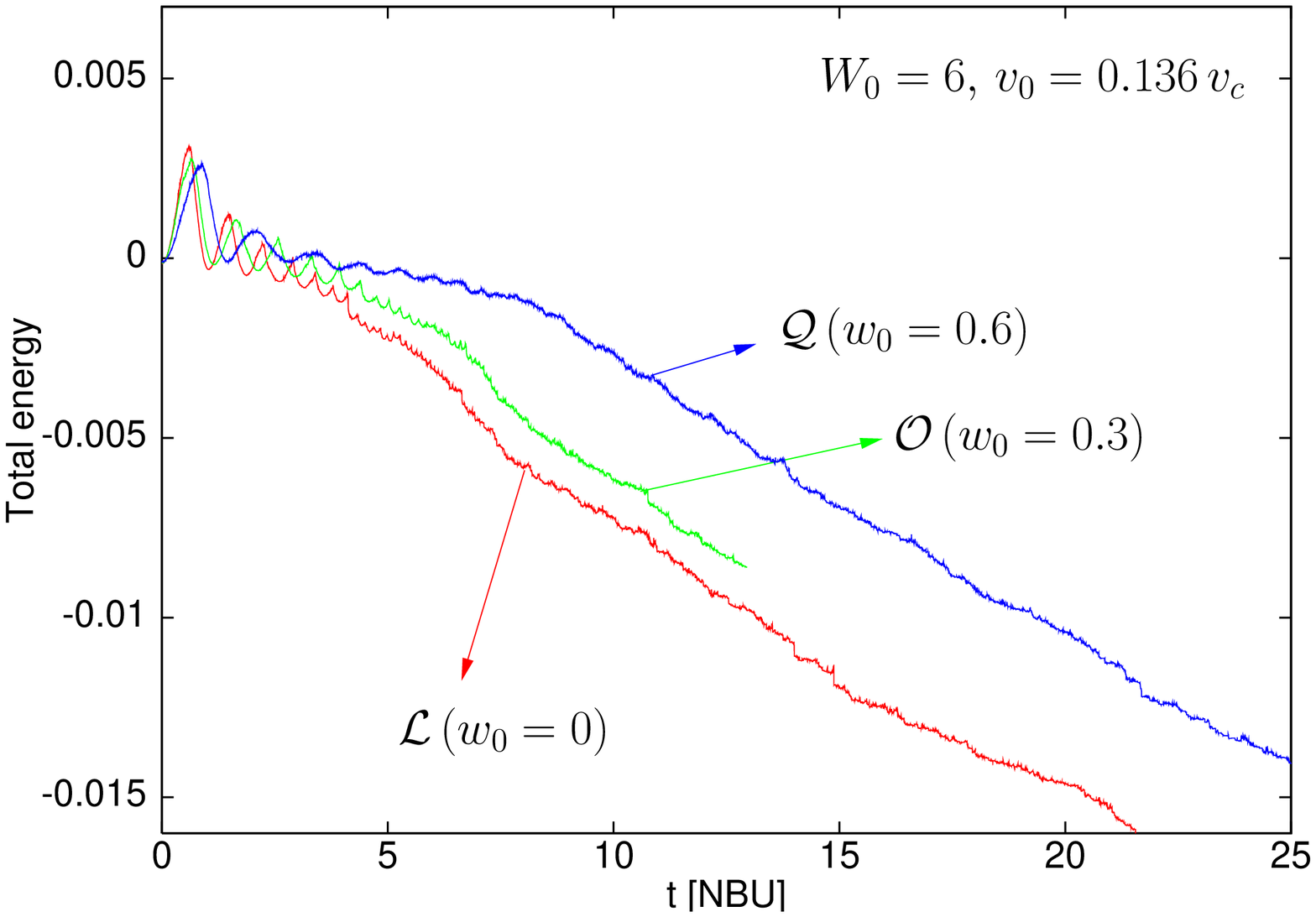}
\caption{
{\em Left panel:} Evolution of the total energy of the IBMH binary for the
models ${\cal I}$, ${\cal F}$ and ${\cal C}$ (from the top to the bottom at later
times; in the on-line version of the paper displayed in blue, green and red
respectively) {\em Right panel:} As in the previous panel, for models ${\cal Q}$,
${\cal O}$ and ${\cal L}$
}\label{energy1}
\end{figure*}

\begin{figure*}
\includegraphics[width=6.0cm,angle=0]{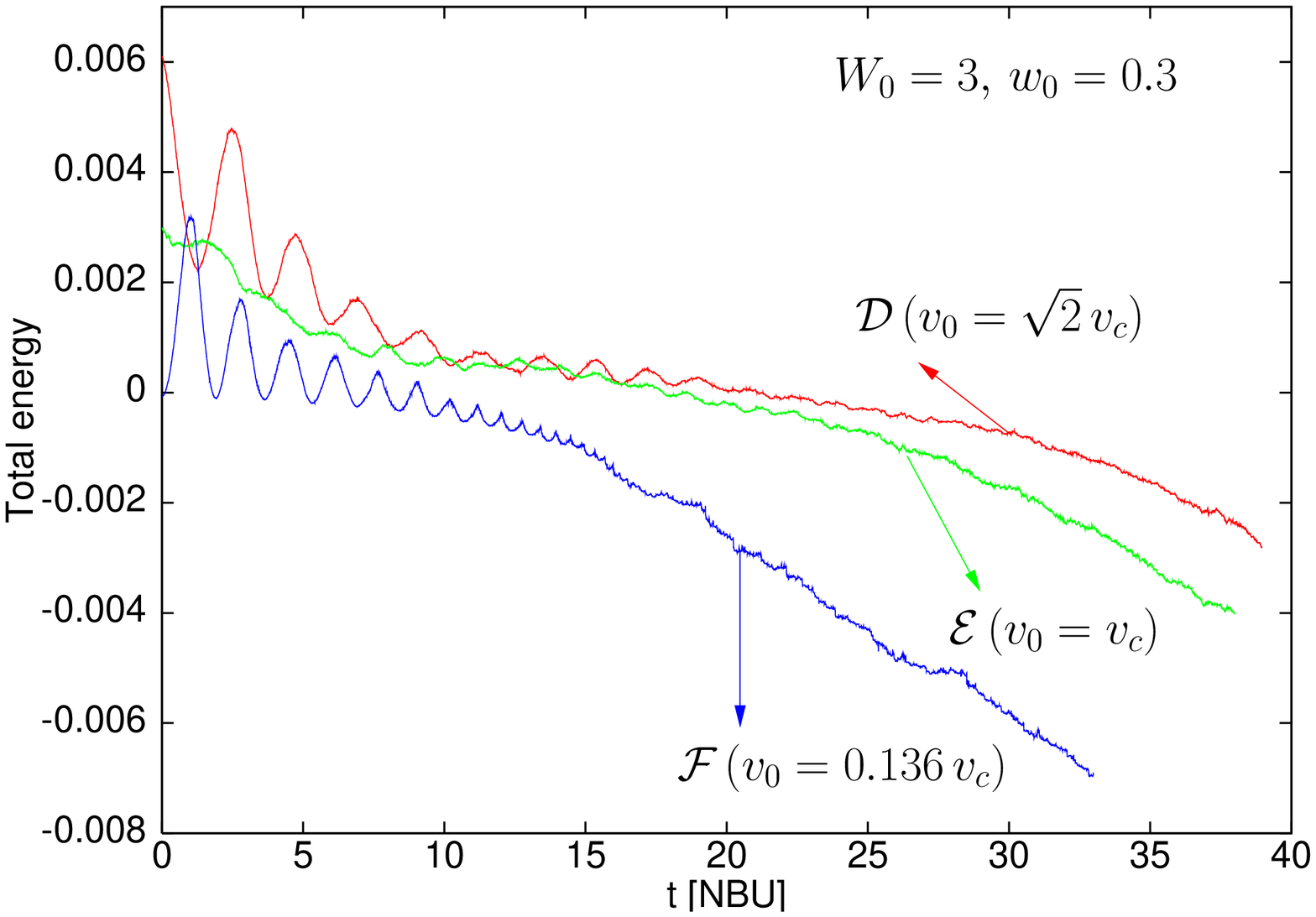}
\includegraphics[width=6.0cm,angle=0]{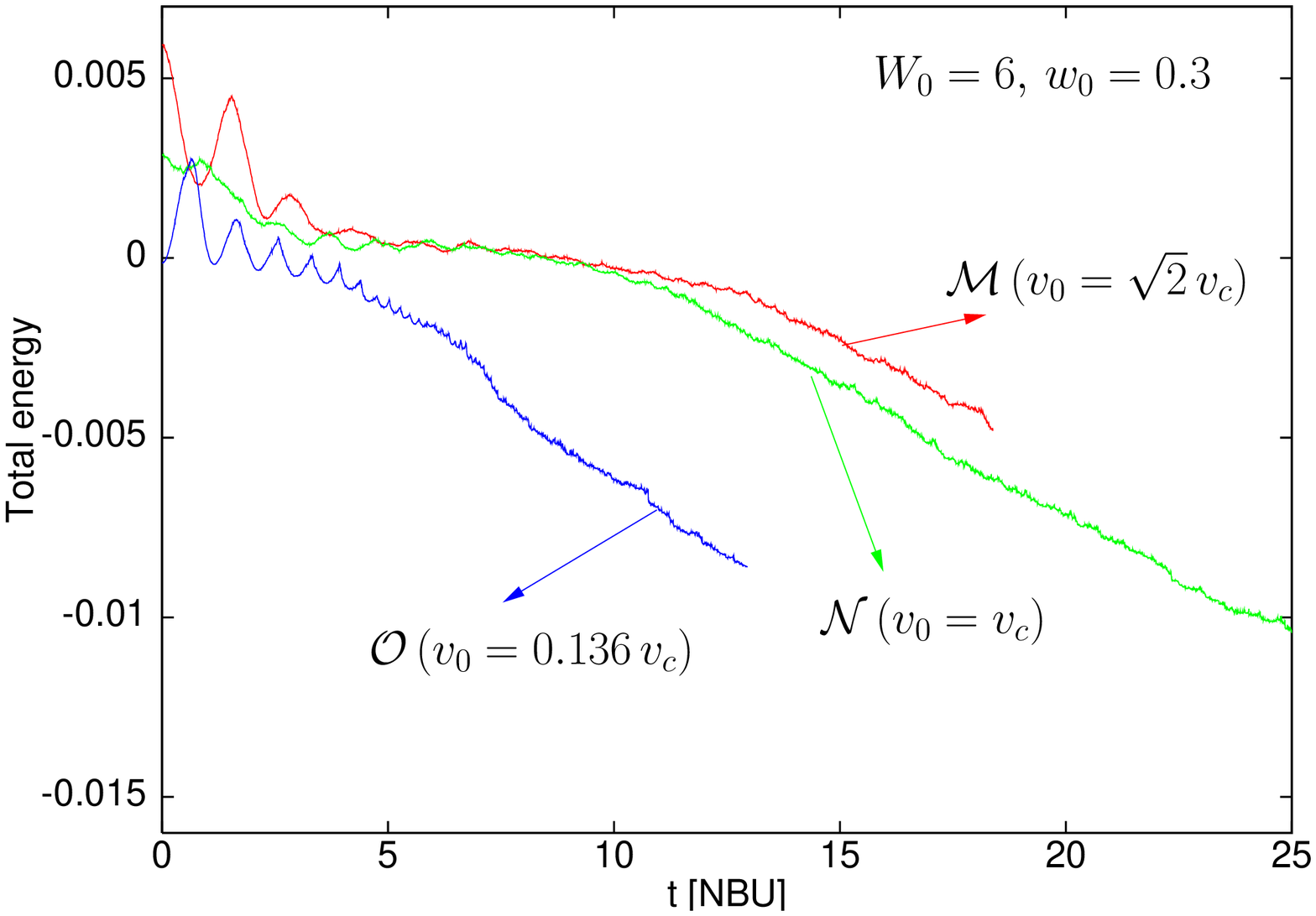}
\caption{
{\em Left panel:} Same as Fig.(\ref{energy1}) for models
 ${\cal D}$, ${\cal E}$ and ${\cal F}$. In this case all models
 have the same King parameter $W_{0}=3$ and $w_0=0.3$, so that
 they differ only in the initial velocities (from the top to the bottom at later times
 $v_{0}=\sqrt{2}v_{c}$, $v_{0}=v_{c}$ and $v_{0}=0.136v_{c}$, depicted in the
 on-line version in red, green and blue, respectively)
{\em Right panel:} As in the previous panel but with $W_{0}=6$, for models 
 ${\cal M}$, ${\cal N}$ and ${\cal O}$
}\label{energy2}
\end{figure*}

The total energy of the binary in a two-body approximation is
given by
\begin{equation}\label{5}
E=\frac{\mu}{2}\dot{r}^{2}+\frac{l^{2}}{2\mu
r^{2}}-\frac{GM_{1}M_{2}}{r}
\end{equation}
where $r$ denotes their separation, $M_{1}$ and $M_{2}$ the black
hole masses, $\mu=M_{1}M_{2}/(M_{1}+M_{2})$ the reduced mass, $l$
the angular momentum and $G$ the gravitational constant.

\begin{table}
\centering
\begin{tabular}{c|c|c|c|c|c}
\hline
${\rm Model}$ & $\sigma_{0}^{2}$ & $\frac{n_{0}}{10^{3}}$ & $\frac{1}{a_{h}}$ & $H$ \\
  \hline
${\cal C}$ & 0.526 & 41.06 & 210 & 7.8$\pm$2.0 \\
  \hline
${\cal F}$ & 0.522 & 43.53 & 209 & 7.2$\pm$1.9 \\
  \hline  
${\cal J}$ & 0.478 & 117.12 & 191 & 5.2$\pm$1.3 \\
${\cal K}$ & 0.478 & 117.12 & 191 & 5.6$\pm$1.5 \\
${\cal L}$ & 0.478 & 117.12 & 191 & 6.1$\pm$1.6 \\
 \hline 
${\cal N}$ & 0.468 & 108.04 & 187 & 5.8$\pm$1.5 \\
${\cal O}$ & 0.468 & 108.04 & 187 & 6.8$\pm$1.7 \\
 \hline 
${\cal Q}$ & 0.425 & 64.79 & 170 & 10.4$\pm$2.7 \\
  \hline
\end{tabular}
\caption{The hardening constants determined in simulations for a
variety of models and initial conditions. $\sigma_{0}$ represents
the initial central velocity dispersion, $n_{0}$ the initial
central particle density, $a_{h}$ the characteristic separation
for hardening via super-elastic scattering processes and $H$ the
hardening constant.} \label{hard}
\end{table}
Fig. \ref{energy1} shows the time evolution of the total energy.
In both cases, the initial velocity is $v_{0}=0.136v_{c}$ on the
core radius, different colours represent different rotational
parameters. Naturally, as long as the gravitational force of the
stellar system dominates the motion of the black holes, the
two-body energy is not very meaningful. Initial oscillations are
the result of this invalidity: Due to the symmetric set-up both
black holes reach the apoapsis and the periapsis almost
simultaneously, in the apoapsis (where their separation is at
maximum, which consequently means a local minimum in the total
energy) the black holes are formally bound to each other ($E<0$).
However, they feel the potential of the stellar system not
included in eq.\ref{5} and are accelerated to the centre while
gathering kinetic energy in such a way that the bound state is
resolved. In this first stage, each black hole individually
suffers dynamical friction, which is the main process of losing
energy.

The role of dynamical friction decreases when a permanently bound
state occurs (the energy remains negative), as the dynamical
friction force acts primarily on the motion of the now formed
binary rather than on the individual black holes. Super-elastic
scattering events of field stars at the binary become more and
more important for the reduction of its energy. These events
become visible in the tiny-peak structure that appears in each
curve in Fig.\ref{energy1} at times $t \gtrsim 17$ for $W_{0}=3$
and $t \gtrsim 8$ for $W_{0}=6$ respectively. In the stage where
super-elastic scattering dominates the picture, the energy loss
rate is commonly written in terms of the dimensionless hardening
constant $H$
\begin{equation}\label{6}
\frac{d}{dt}\left(\frac{1}{a}\right)= H G\frac{\rho}{\sigma}
\end{equation}
where $a$ is the separation of the black holes, $\rho$ the mass
density and $\sigma$ the velocity dispersion in the environment of
the binary \citep[see e.g.][]{mer2001}. The constant slope of the
energy in Fig.\ref{energy1} is expected from eq.\ref{6} with
$\rho/\sigma=$const.

In Tab.\ref{hard}, hardening constants have been determined for scenarios in
which a constant energy loss rate had developed before the simulation ended.
The time derivative of the inverse separation was taken from the slope of the
curves in Fig.\ref{energy1}, which is connected to the energy by $E=-G\mu/2a$.
Approximately, $\sigma/\rho=\sigma_{0}/\rho_{0}$ was assumed with $\rho_{0}$
and $\sigma_{0}$ as initial values within the 1\% Lagrangian radius of the
model. As this represents a rather vague approximation, a 25\% range of this
ratio was combined with an regression estimate of the uncertainties in slope
designation to obtain the error margins. 

Since the stage which is dominated by superelastic scattering is reached sooner
if the central potential of the stellar distribution is deeper, the hardening
constants have been calculated primarily for runs with $W_0=6$ in Table 2.  For
these runs, the separation of the IMBHs has fallen below the characteristic
separation $a_h=GM/4\sigma^2_0$ at the end of the simulation, which indicates
that the system is in the hardening regime.

The values of the hardening constants are in majority slightly
below compared to the $H=8.4$ published by \citet{hem2002}, where
a Plummer model was used. The lower values can be possibly
explained by the fact that dynamical friction might still have a
noticeable influence. Regarding our calculated $a_{h}$ as criteria
for the domination of super-elastic scattering events, the
hardening separation could be significantly smaller if $\sigma$
increases during the simulation as $a_{h}\propto\sigma^{-2}$. An
enhanced $\sigma$ can be expected for $\rho/\sigma=$const. if it
is assumed that the black hole would capture stars during the
simulation and raise the central density.

We can see in Fig. 1 that King potentials of $W_0=3$ as well as $W_0=6$ the
two-body energy reaches higher values (i.e. the system is less bound) at a
given time for $\omega_0=0.6$, as compared to the simulations with
$\omega_0=0.0$ and $\omega_0=0.3$. Additionally, we find that in the presence
of faster rotation $\omega_0=0.6$, the transfer of angular momentum to the
field stars is inhibited during the first 2--4 time units. This can lead to the
circularising of the IMBHs trajectories and, indeed, we find remarkable smaller
eccentricities for this case (see Fig 6 of next section).

Variation of the initial velocities is presented in
Fig.\ref{energy2}. In simulations with $v_{0}=v_{c}$, no
oscillations occur in the first time units, as the black holes
spiral to the centre symmetrically on circular-like orbits.

\subsection{Eccentricity}
The eccentricity is determined by
\begin{equation}\label{651}
e=\sqrt{1+\frac{2El^{2}}{\mu (GM_{1}M_{2})^{2}}}
\end{equation}

\begin{figure*}
\subfigure{\includegraphics[width=6.0cm,angle=0]{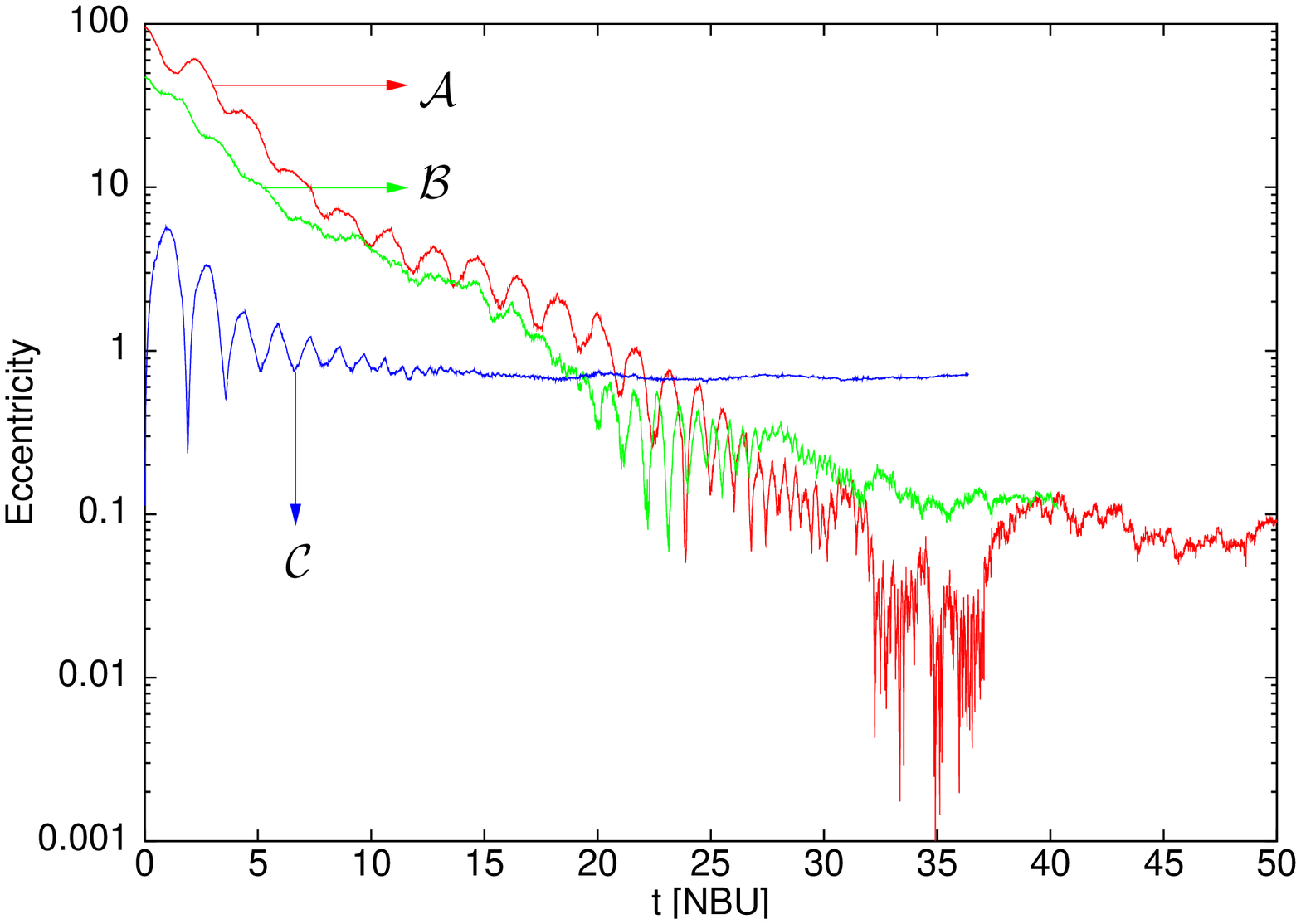}
\includegraphics[width=6.0cm,angle=0]{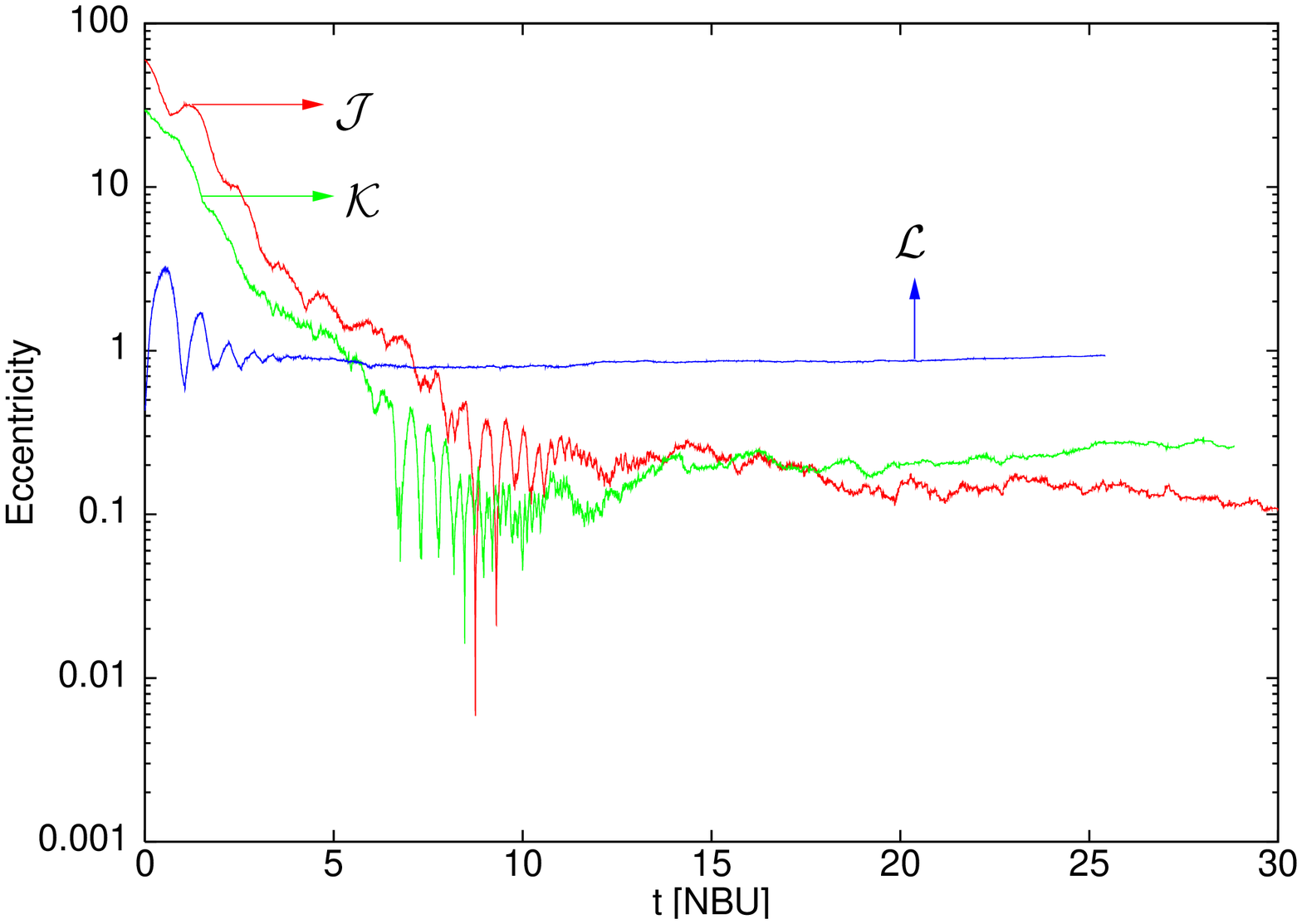}}\\
\subfigure{\includegraphics[width=6.0cm,angle=0]{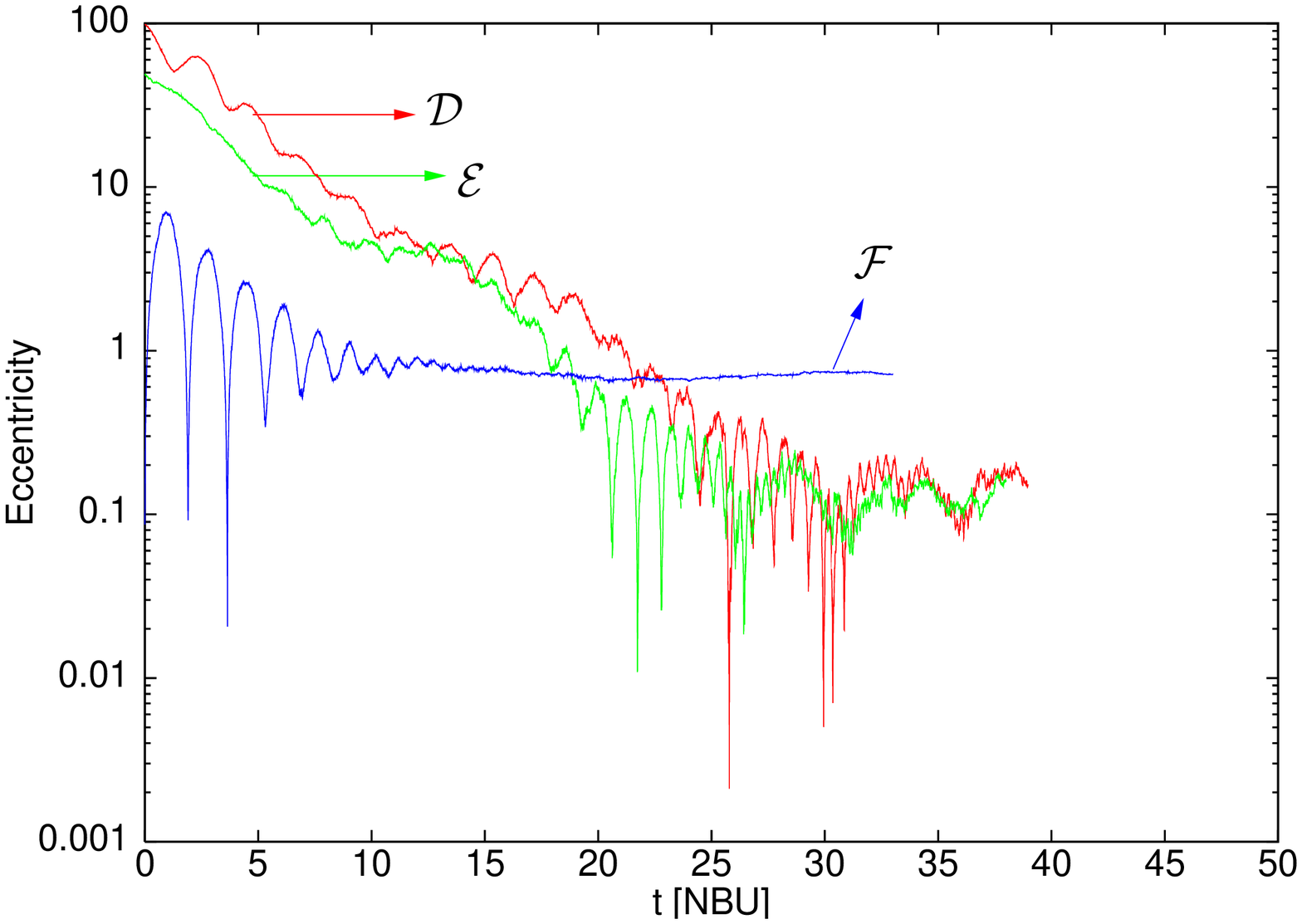}
\includegraphics[width=6.0cm,angle=0]{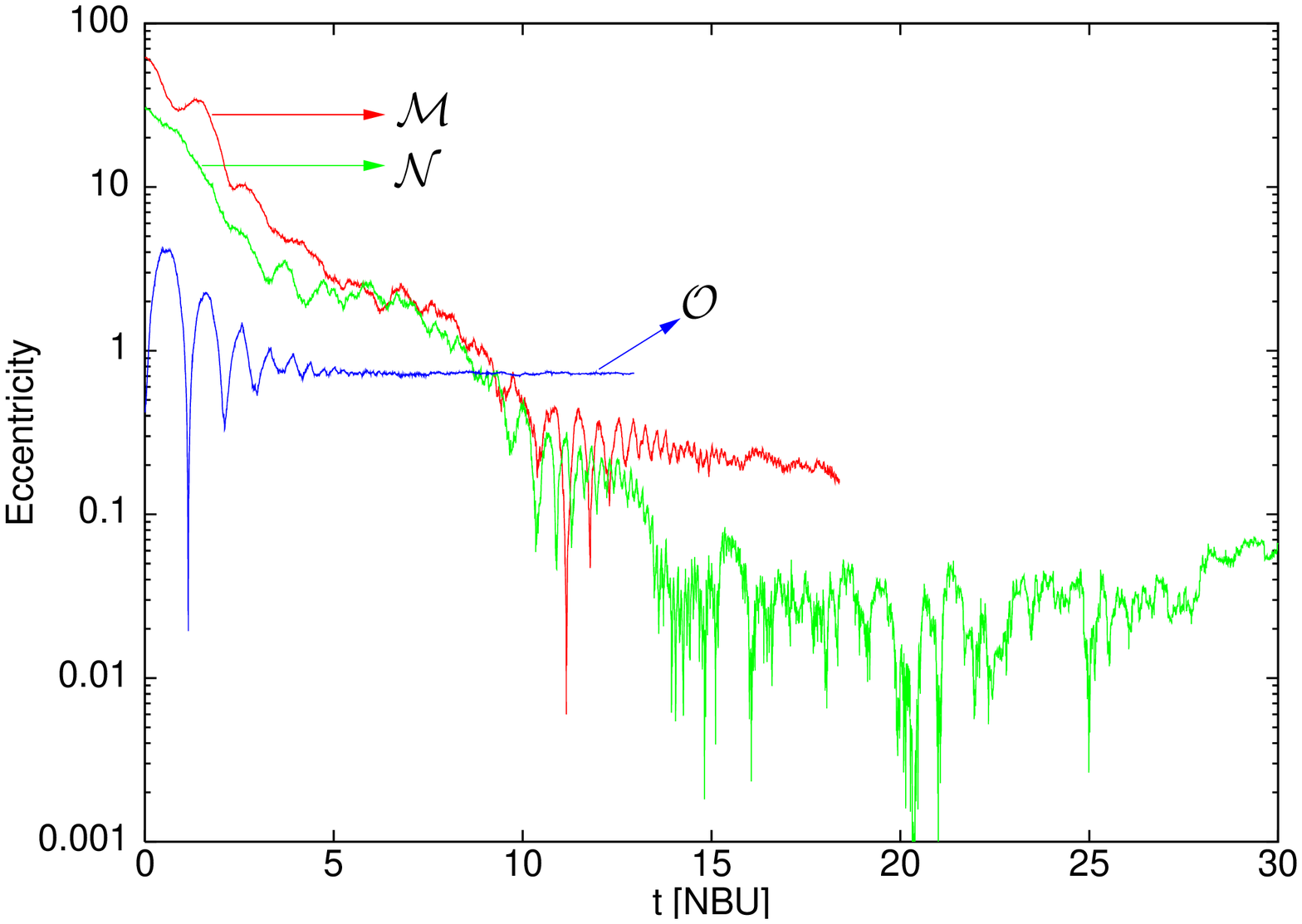}}\\
\subfigure{\includegraphics[width=6.0cm,angle=0]{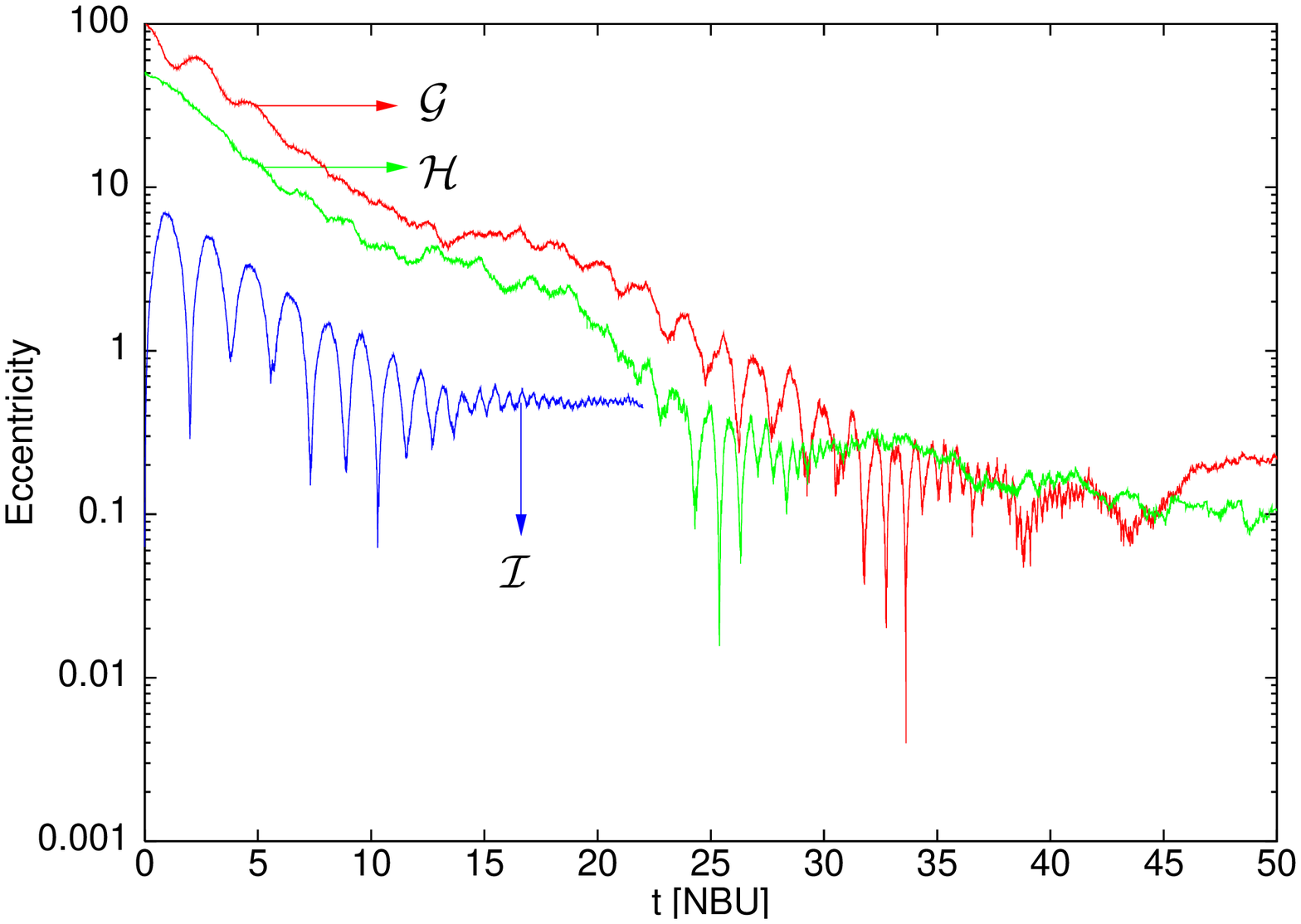}
\includegraphics[width=6.0cm,angle=0]{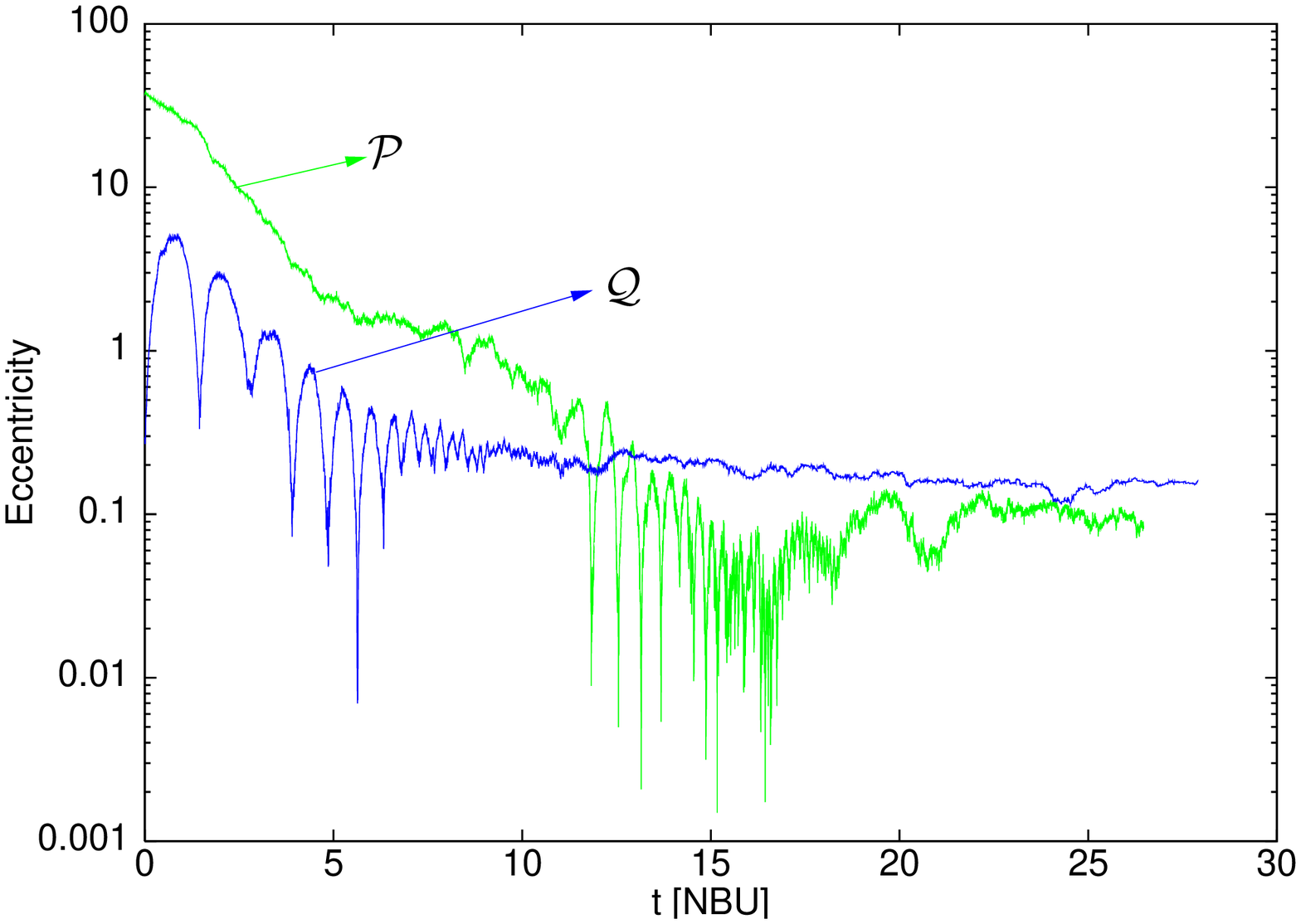}}
\caption{
Evolution of the eccentricity of the IMBH binary for the complete
survey. Starting from the top and from the left to the right,
we display in each panel only three models for readability. In each
individual panel, and from the top to the bottom at $t = 0$, we show
(in red, green and blue in the on-line version):
${\cal A}$, ${\cal B}$, ${\cal C}$ in the first panel, ${\cal J}$, ${\cal K}$,
${\cal L}$ in the second one, ${\cal D}$, ${\cal E}$, ${\cal F}$ in the third
one, ${\cal M}$, ${\cal N}$, ${\cal O}$ in the fourth one, ${\cal G}$, ${\cal
H}$, ${\cal I}$ in the fifth one and ${\cal P}$, ${\cal Q}$ in the last one
}\label{allecc}
\end{figure*}

Fig.\ref{allecc} shows the complete survey over calculated
eccentricity evolution. Each plot shows simulations of a fixed
pair of King parameters under variation of the initial velocity.
In the case of $W_{0}=6$, $\omega_{0}=0.6$, no simulation could be
performed with $v_{0}=\sqrt{2}v_{c}$ without overstepping the
error limits mentioned in section \ref{sec.initial} holding the set of NBODY
accuracy parameters. The runs were stopped when a fixed physical
calculation time of the PC cluster was exceeded.

Initial oscillations appear for the same reasons as in the plots
of the total energy previously discussed. After the binary has been formed, and
in principle represents a two body system perturbed by
encountering field stars, the eccentricity converges to a fixed
value that underlies at most a weak drift. This behaviour is
consistent with previous work \citep{hem2000,hem2002}. When the
eccentricity has swung into a certain level, again a tiny-peak
structure develops as the result of super-elastic scattering
processes of field stars at the hardening binary. Note that
following the swing-in-procedure, the stochastic fluctuations
are of the same order, due to the logarithmic scaling.

All simulations displayed in Figure \ref{allecc} with an initial velocity
comparable to the circular velocity, tend to end up in low-eccentricity motions
of the black hole components, while $v_{0}=0.136v_{c}$ runs reach generally
higher final eccentricities. This behaviour was already indicated by
\citet{mak1993}, who simulated two black holes of the masses $M=0.01$ in a
Plummer sphere of 16348 particles. They found very high final eccentricities
$e\sim0.99$ applying very low initial velocities, while their largest value,
$v_{0}=0.5v_{c}$, reached a noticeably smaller final $e\sim 0.665$.
Investigations carried out by \citet{hem2000} and \citet{hem2002} used initial
velocities $v_{0}=0.136v_{c}$, their high eccentricities were verified in the
simulations presented here for the rotating King model.

\begin{figure*}
\subfigure{\includegraphics[bb= 60 55 570
550,clip,scale=0.50,angle=270]{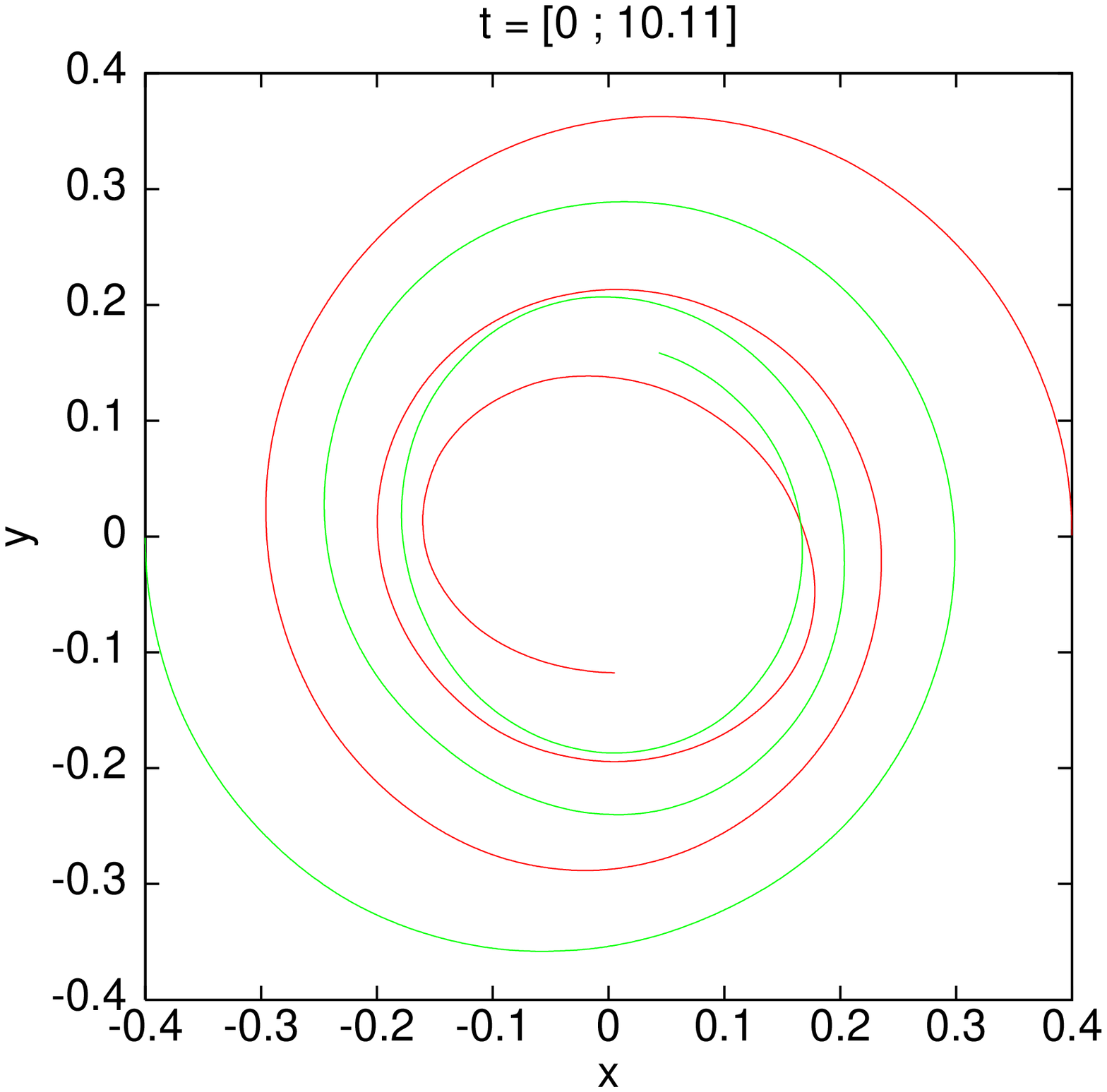}
\includegraphics[bb= 60 60 570
550,clip,scale=0.50,angle=270]{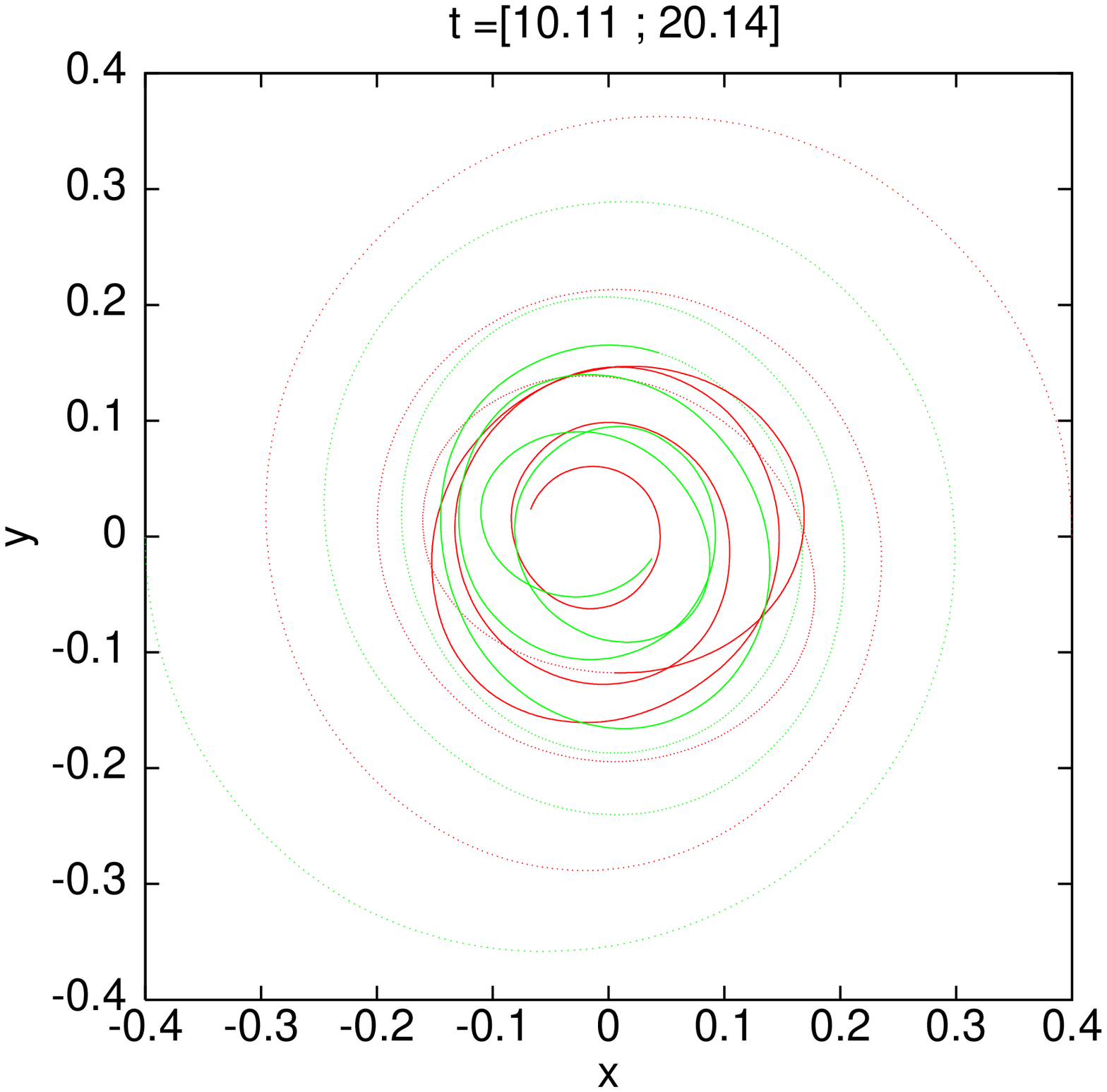}}\\
\subfigure{\includegraphics[bb= 60 60 570
550,clip,scale=0.50,angle=270]{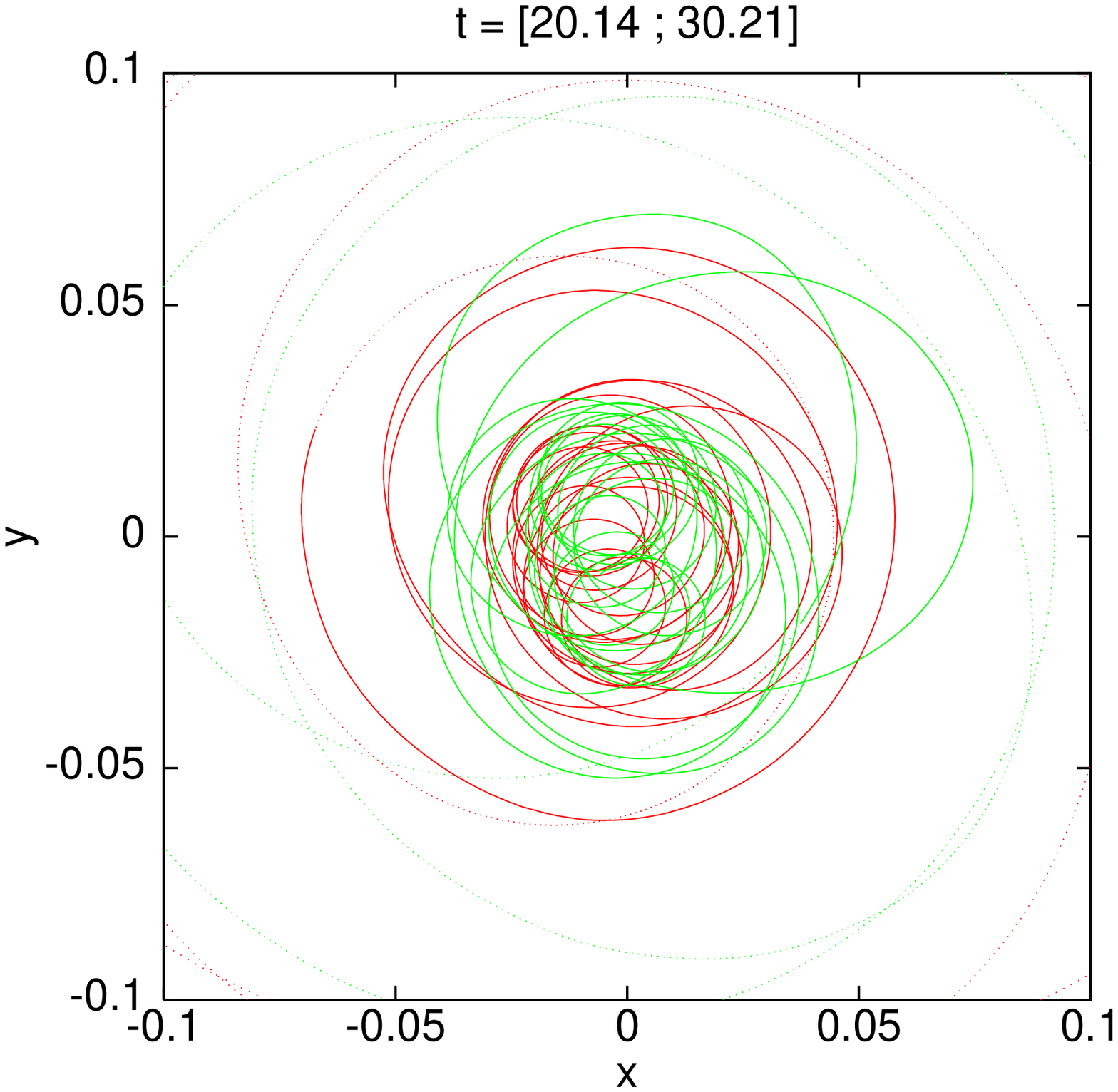}
\includegraphics[bb= 60 60 570 550,clip,scale=0.50,angle=270]{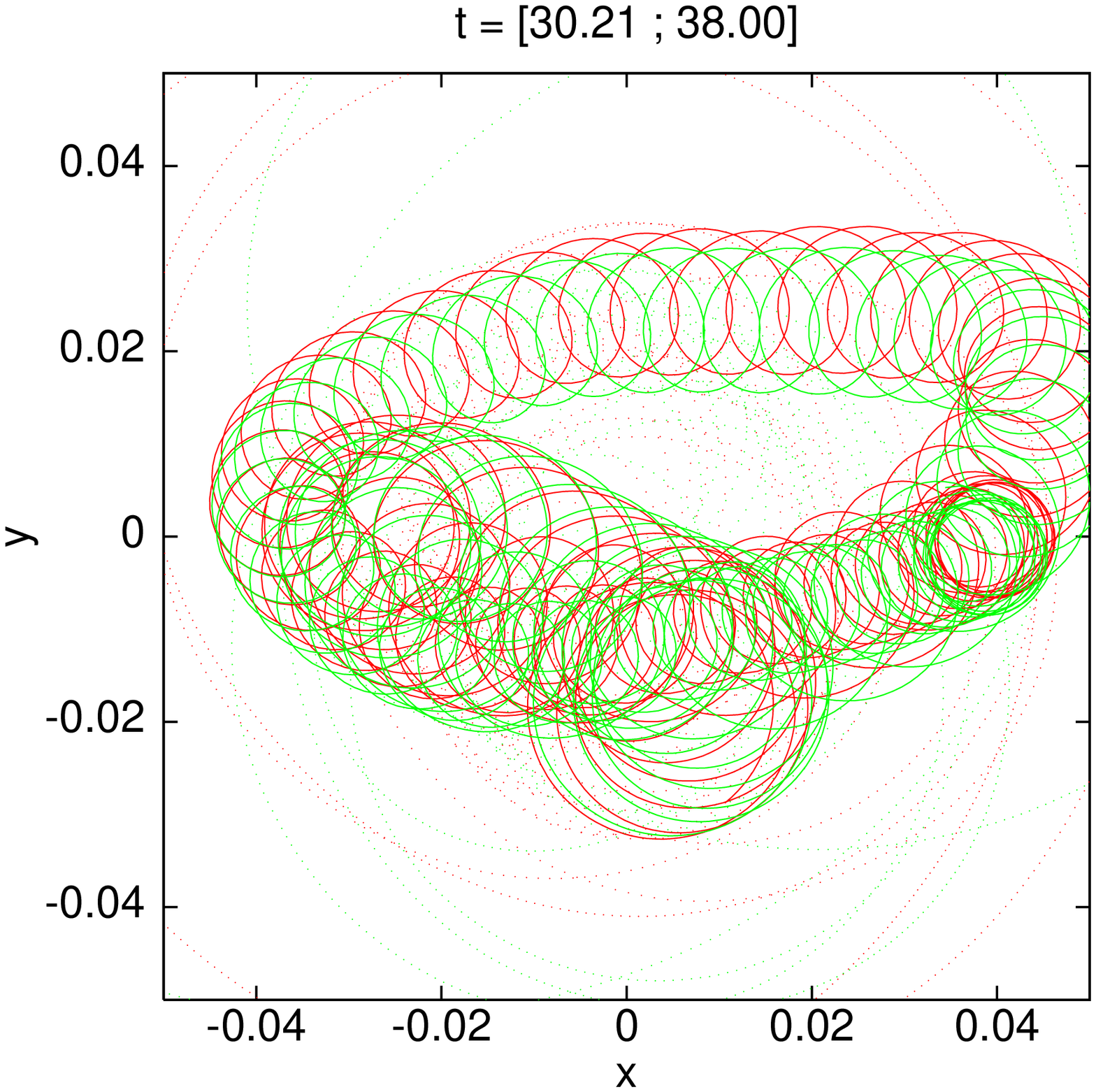}}
\caption{
Projection in the X--Y plane of the trajectories of the two IMBHs
in the model ${\cal E}$ (different colours depict different IMBHs).
Solid lines indicate the trajectories passed through in the time
interval mentioned above each figure; the dotted lines
show the previous orbit. Note the scaling of the axes in different figures;
the two lower panels are a zoom in
}
\label{traj1}
\end{figure*}

The dependency of the final eccentricity on initial velocities can
be understood by considering the black hole trajectories. In Fig.
\ref{traj1}, for $v_{0}=v_{c}$, the black holes spiral, at first
independently of each other, to the centre. The influence of
dynamical friction causes a steady loss of kinetic energy. Within
the time interval $t=[10.11;20.14]$, the total energy becomes
negative and the binary reaches a bound state; subsequently
the binary hardens, the separation decreases due to super-elastic
scattering events and the circular motion of the centre of mass of the
binary itself becomes visible. At the time the attractive force
between the black holes becomes comparable to the gravitational
force of the stellar distribution, the individual trajectories of
the black holes are still circular around the systems centre of
mass. This means that the circular orbits generated by the initial
velocity is "conserved" until the binary reaches a bound state and
beyond, since dynamical friction is not strong enough to change the
trajectories dramatically.

\begin{figure*}
\subfigure{\includegraphics[bb= 60 55 570
550,clip,scale=0.50,angle=270]{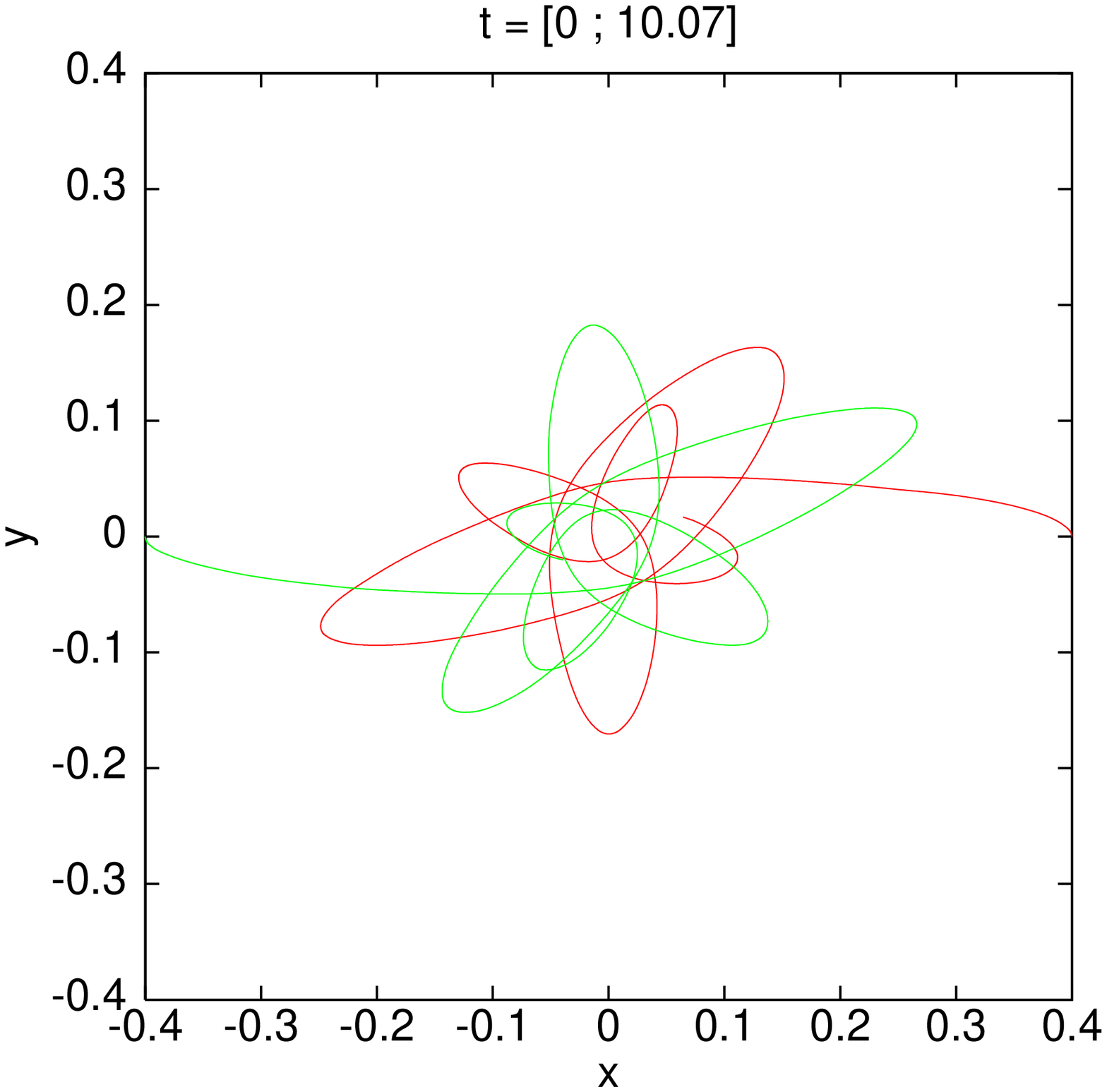}
\includegraphics[bb= 60 60 570
550,clip,scale=0.50,angle=270]{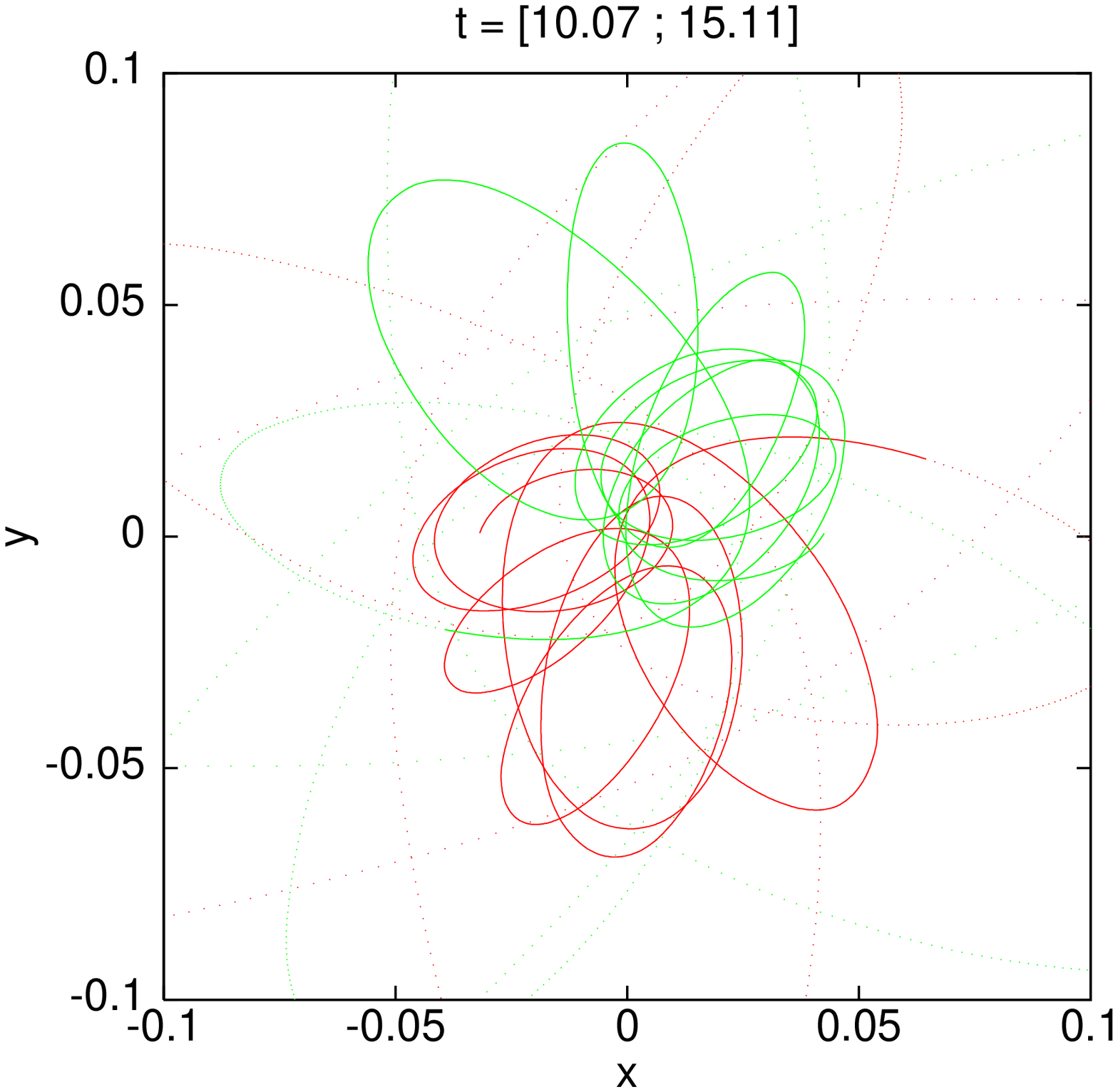}}
\subfigure{\includegraphics[bb= 60 60 570
550,clip,scale=0.50,angle=270]{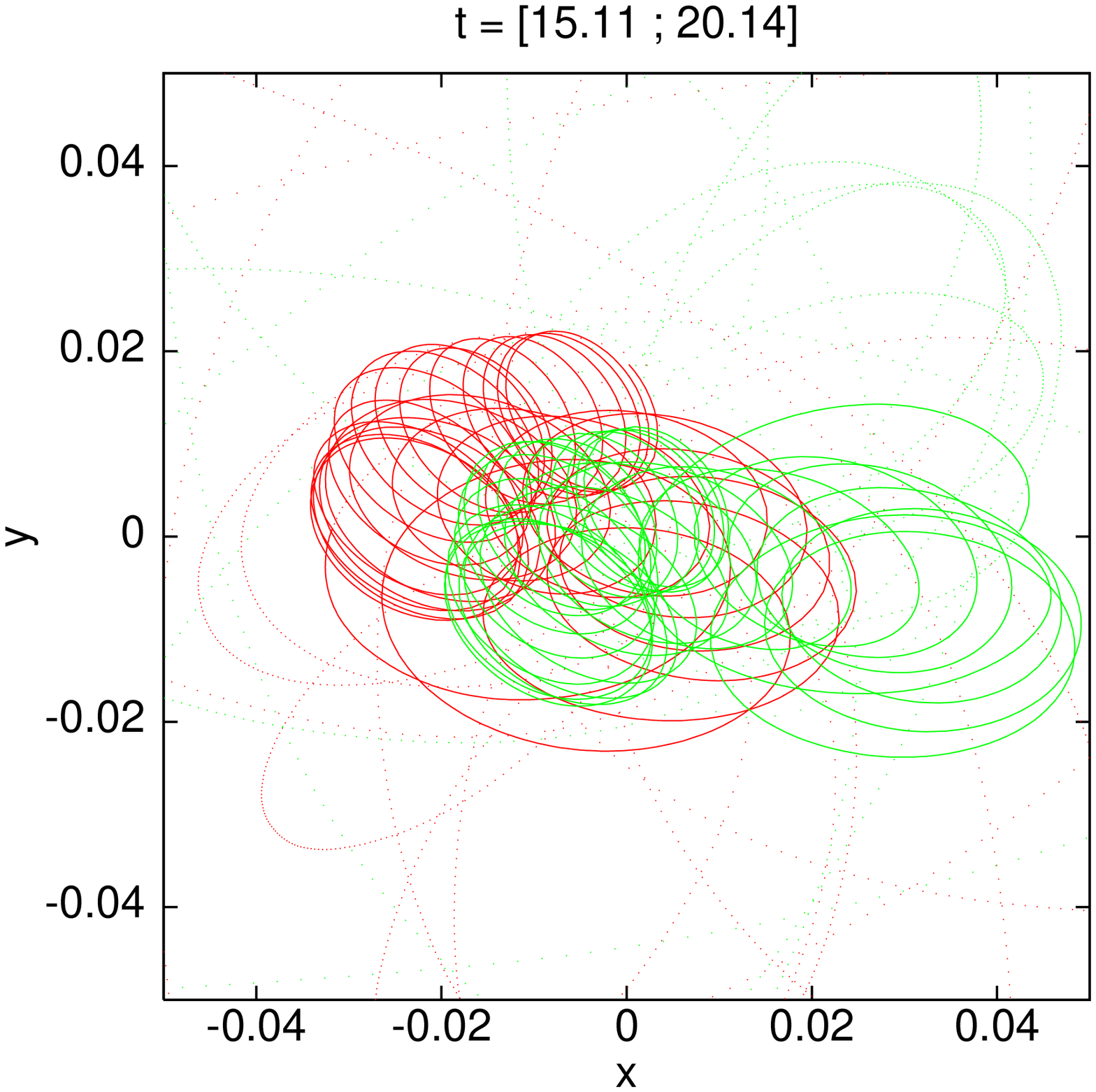}
\includegraphics[bb= 60 60 570 550,clip,scale=0.50,angle=270]{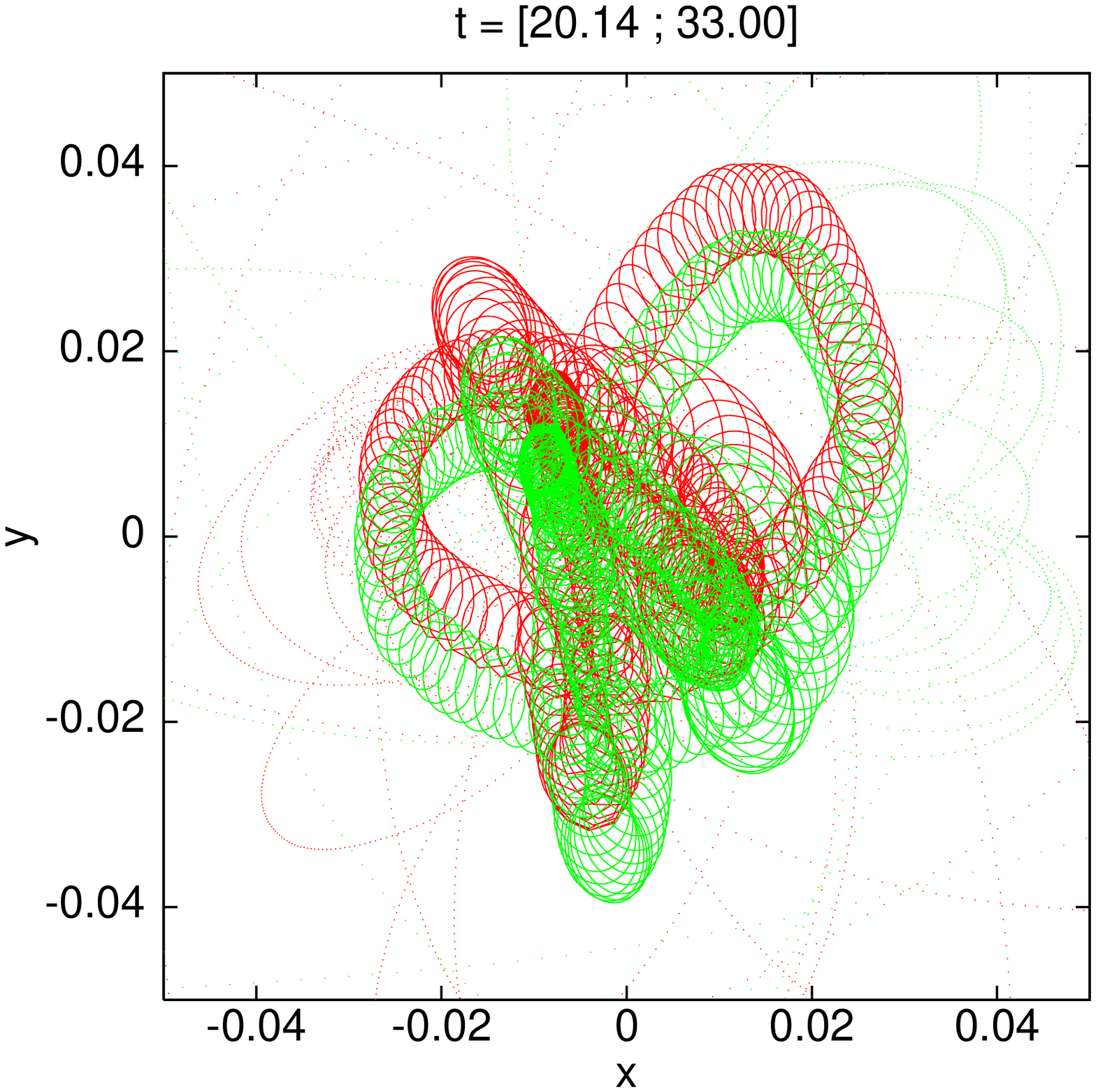}}
\caption{the same as in Fig.\ref{traj1}, but for model ${\cal F}$
}\label{traj2}
\end{figure*}

A different situation is obtained for $v_{0}=0.136\,v_c$. As a
consequence of the low velocity, the black holes must plunge near
to the centre, but dynamical friction is, at the time of the
closest encounter (the periapsis of the relative motion), not
sufficiently effective to prevent the re-swing to the outer regions
and to circularise the orbits in this way. Therefore, the initial
form of the orbits is kept until the end of the simulation.
The initial velocity $v_{0}=\sqrt{2}\,v_c$ is also non-circular.
The deviations from the previous case are due to the fact that 
dynamical friction is stronger at apoapsis.

Tab. \ref{ecctab} summarises the results of the complete study.
The mean values of the eccentricity $\bar{e}$, the error of the
single value $s_{e}$ and the error of the mean value $s_{\bar{e}}$
have been calculated over time intervals in which the eccentricity
remains stable, and are based on output data of a step-width of
0.002 N-body units. The quantity $e_{\rm trans}$ represents a mean
value of the eccentricity in a period the total energy snaps of
into the stage of constant energy loss. Although the the
eccentricity still undergoes vigorous fluctuations at that time,
this transition mean value $e_{\rm trans}$ does not evidently differ
from the $\bar{e}$. It is remarkable that eccentricity develops in
a dynamical friction dominated period. The strength of the
dynamical friction force along the trajectory of the black holes
is crucial whether the individual motion determined by the initial
velocities can be kept until the binary forms.

The existence of drifts in eccentricity was reported in previous papers (see
introduction). This behaviour can be connected to the effect of super-elastic
scattering events.  Although the effect of super-elastic scatterings on the
eccentricity during the formation of the binary may be only weak, the long term
evolution might increase the eccentricity significantly. However, the question,
whether the eccentricity is ultimately increased or decreased, cannot be solved
here since the simulations includes both rising and declining drifts, as seen
for $W_{0}=6$, $\omega_{0}=0.0$ with $v_{0}=v_{c}$ and $v_{0}=\sqrt{2}v_{c}$
(Fig.\ref{allecc}, for example. If superelastic scattering should actually
increase the eccentricity in long-term evolution, another effect overlays the
simulations in that period.

\begin{figure*}
\includegraphics[width=6.0cm,angle=270]{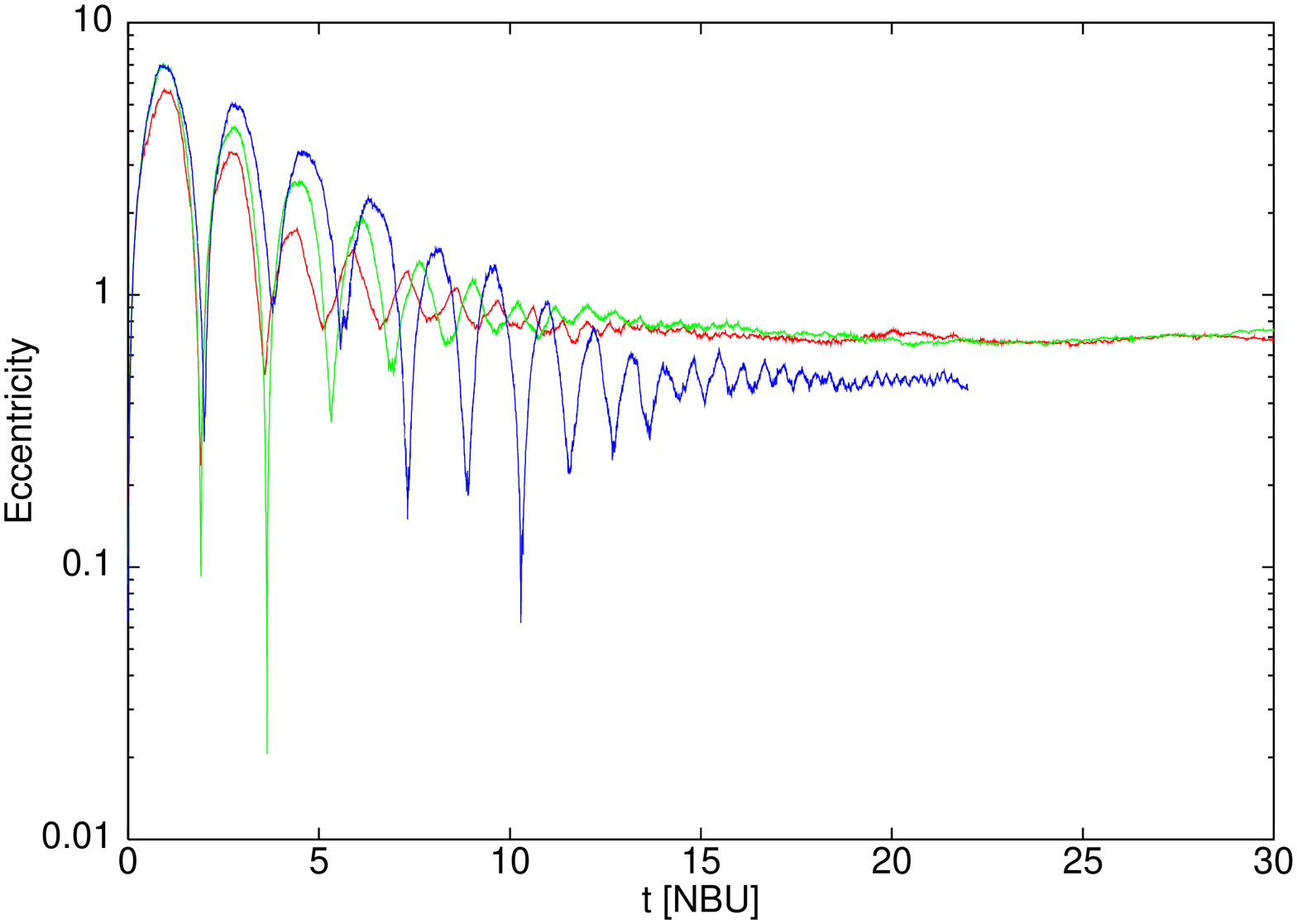}
\includegraphics[width=6.0cm,angle=270]{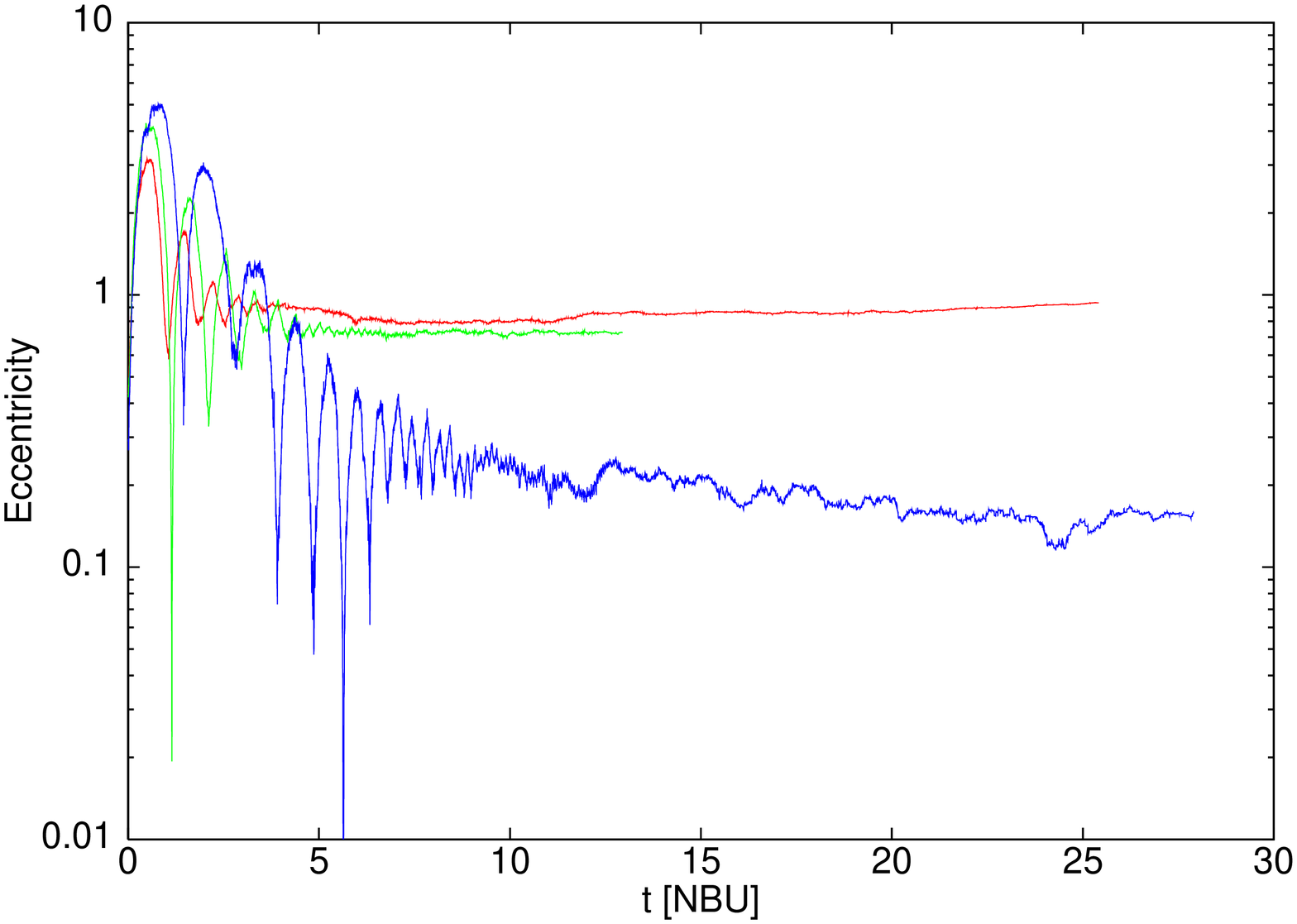}
\caption{
Influence of rotation on final eccentricities.  {\em Left panel: } King
$W_{0}=3$, models ${\cal F}$, ${\cal C}$ and ${\cal I}$ in green, red and blue,
respectively.  {\em Right panel: } King $W_{0}=6$, models ${\cal O}$, ${\cal
L}$ and ${\cal Q}$ in green, red and blue, respectively. The colours indicate
the different rotation parameters used in the simulations: $\omega_{0}=0.0$
\emph{red}, $\omega_{0}=0.3$ \emph{green}, $\omega_{0}=0.6$ \emph{blue}
} \label{Ecc000306}
\end{figure*}

We depict in Fig. \ref{Ecc000306} the cases ${\cal F}$, ${\cal C}$, ${\cal I}$,
${\cal O}$, ${\cal L}$ and ${\cal Q}$ of Fig. \ref{allecc} together in order to
illustrate clearly the influence of rotation on the development of
eccentricity.  With a rotation parameter $\omega_{0}=0.6$, the end value of
eccentricity drops significantly for $W_{0}=3$ as well as for $W_{0}=6$. In the
case of $W_{0}=3$, $\omega_{0}=0.6$, no effective transfer of angular momentum
to the field stars takes place until $t\sim 5$, with $W_{0}=6$,
$\omega_{0}=0.6$ till $t\sim 2$, thus the trajectories of the IMBHs  experience
a noticeable truncation in the co-rotating stellar system at the beginning,
which finally results in low eccentricities.

\begin{table*}
\centering
\begin{tabular}{c|c|c|l|c|c|c|l}
  \hline
  Set & Model & $t_{\rm stab}$ &\qquad \qquad $\bar{\rm e}$
  & $e_{\rm end}$ & $s_{\bar{\rm e}}$ & $s_{\rm e}$ &\qquad $\bar{\rm e}_{\rm trans}$ \\
  \hline
   & ${\cal A}$ & 40 & 0.0812
  [41.00;52.15] & 0.0930 & 0.0002 & 0.0153 & 0.13 [26.4;32.7] \\
 1 & ${\cal B}$ & 34 & 0.1212
  [35.00;40.33] & 0.1069 & 0.0002 & 0.0102 & 0.29 [25.1;28.8] \\
   & ${\cal C}$ & 13 & 0.7272
  [35.00;36.73] & 0.7146 & 0.0002 & 0.0047 & 0.72 [14.7;18.0] \\
  \hline
   & ${\cal D}$ & ? & 0.1447
  [35.00;38.96] & 0.1512 & 0.0008 & 0.0348 & 0.14 [28.2;34.7] \\
 2 & ${\cal E}$ & 35? &0.1239
  [35.00;38.00] & 0.1586 & 0.0005 & 0.0197 & 0.12 [24.9;32.4] \\
   & ${\cal F}$ & 15 & 0.7135 [33.00]
  & 0.7135 & - & - & 0.72 [13.3;16.9] \\
  \hline
   & ${\cal G}$ & 46? & 0.1714
  [40.00;56.00] & 0.2204 & 0.0005 & 0.0462 & 0.15 [32.8;42.0] \\
 3 & ${\cal H}$ & 38 & 0.1206
  [38.00;53.15] & 0.1184 & 0.0003 & 0.0261 & 0.14 [23.0;29.4] \\
   & ${\cal I}$ & 19 & 0.4497 [22.00]
  & 0.4497 & - & - & 0.50 [14.9;18.9] \\
  \hline
   & ${\cal J}$ & 15 & 0.1369
  [22.00;30.46] & 0.1349 & 0.0003 & 0.0170 & 0.25 [8.3;11.6] \\
 4 & ${\cal K}$ & 14 & 0.2517
  [22.00;28.84] & 0.2607 & 0.0004 & 0.0228 & 0.13 [7.1;9.4] \\
   & ${\cal L}$ & 3 & 0.9116
  [22.00;25.41] & 0.9359 & 0.0003 & 0.0127 & 0.90 [5.7;7.6] \\
  \hline
   & ${\cal M}$ & ? & 0.1558
  [18.38] & 0.1558 & - & - & 0.26 [12.1;14.1] \\
 5 & ${\cal N}$ & 16 & 0.0387
  [22.00;30.58] & 0.0878 & 0.0003 & 0.0191 & 0.19 [8.8;11.6] \\
   & ${\cal O}$ & 5 & 0.7276 [12.95]
  & 0.7276 & - & - & 0.71 [5.8;7.0] \\
  \hline
   & ${\cal P}$ & 22 & 0.0914
  [25.00;26.47] & 0.0813 & 0.0003 & 0.0073 & 0.07 [12.1;16.0] \\
 6 & ${\cal Q}$ & 13 & 0.1534
  [25.00;27.90] & 0.1599 & 0.0002 & 0.0075 & 0.25 [6.7;8.3] \\
  \hline
\end{tabular}
\caption{Compiled data of eccentricity evolution for the the
complete set of simulations. $t_{\rm stab}$ represents the time when the
eccentricity remains passably constant, $\bar{\rm e}$ is the average
eccentricity measured over the time interval mentioned below,
${\rm e}_{\rm end}$ the value at the end of the simulation. Disregarding
appearing drifts, errors were calculated based on an 0.002
step-width output considering the same time intervals; ${\rm s}_{\rm e}$ is
the error of the single value and $\bar{\rm e}_{\rm trans}$ the error of
the mean value eccentricity. $\bar{\rm e}_{\rm trans}$ is an average
eccentricity over the epoch when the system turns to a constant
energy loss rate}\label{ecctab}
\end{table*}

\subsection{Inclination and angular momentum orientation}
\begin{table}
\begin{tabular}{c|c|c}
  \hline
  Model & ${\rm i}_{\rm max}$ & $\bar{\rm i}$\\
  \hline
  ${\cal A}$ & 0.191 [37.151] & 0.14 [32.7;52.2]\\
  ${\cal B}$ & 0.360 [19.275] & 0.27 [28.8;40.3]\\
  ${\cal C}$ & 0.362 [3.316]  & 0.23 [18.0;36.7]\\
  \hline
 ${\cal D}$ & 0.271 [18.557] & 0.04 [34.7;39.0]\\
 ${\cal E}$ & 0.289 [12.836] & 0.21 [32.4;38.0]\\
 ${\cal F}$ & 0.352 [3.329]  & 0.07 [16.9;33.0]\\
  \hline
 ${\cal G}$ & 0.189 [24.961] & 0.14 [42.0;56.0]\\
 ${\cal H}$ & 0.201 [23.056] & 0.13 [29.4;53.2]\\
 ${\cal I}$ & 0.344 [1.686]  & 0.12 [18.9;22.0]\\
  \hline
 ${\cal J}$ & 0.316 [15.869] & 0.24 [11.6;30.5]d\\
 ${\cal K}$ & 0.252 [7.138]  & 0.18 [9.4;28.8]\\
 ${\cal L}$ & 0.867 [20.895] & 0.72 [7.6;25.4]\\
  \hline
 ${\cal M}$ & 0.349 [11.120] & 0.24 [14.1;18.4]\\
 ${\cal N}$ & 0.178 [9.546]  & 0.07 [11.6;30.6]\\
 ${\cal O}$ & 0.492 [5.202]  & 0.42 [7.0;13.0]\\
  \hline
 ${\cal P}$ & 0.184 [26.423] & 0.15 [16.0;26.5]\\
 ${\cal Q}$ & 0.158 [10.150] & 0.09 [8.3;27.9]d\\
  \hline
\end{tabular}
\caption{Inclination characteristics regarding all simulations.
${\rm i}_{\rm max}$ denotes the maximum value at the mentioned time unit,
$\bar{\rm i}$ is the mean value over the mentioned time interval,
which ranges from the time unit when a constant energy loss rate
occurs until the end of the simulation. The mean values are
calculated despite some runs showing drift, in which case they are indicated 
with a ''d''.}\label{inctab}
\end{table}

\begin{figure*}
\includegraphics[width=7.2cm,angle=270]{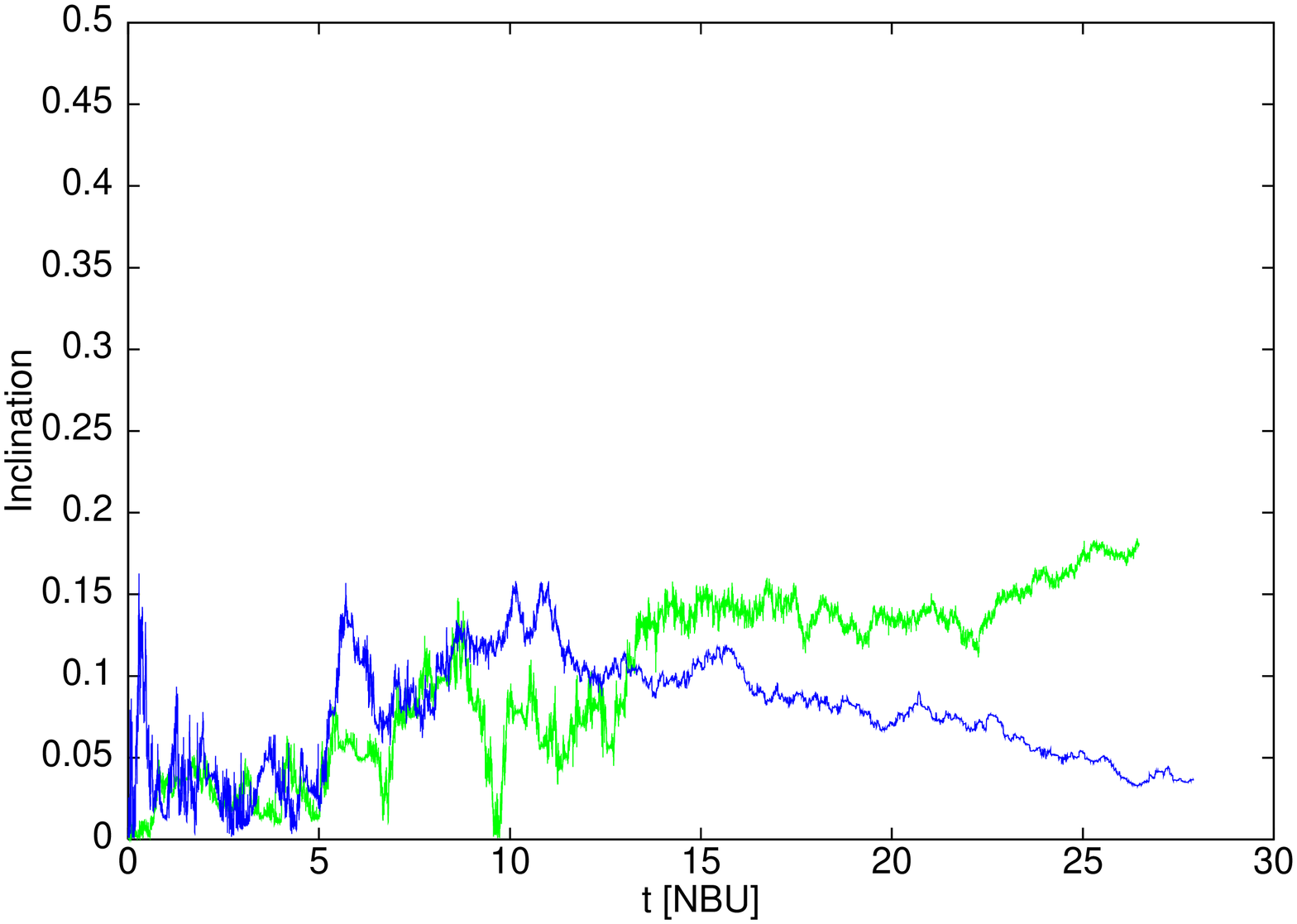}
\includegraphics[bb= 60 55 570
550,clip,scale=0.38,angle=270]{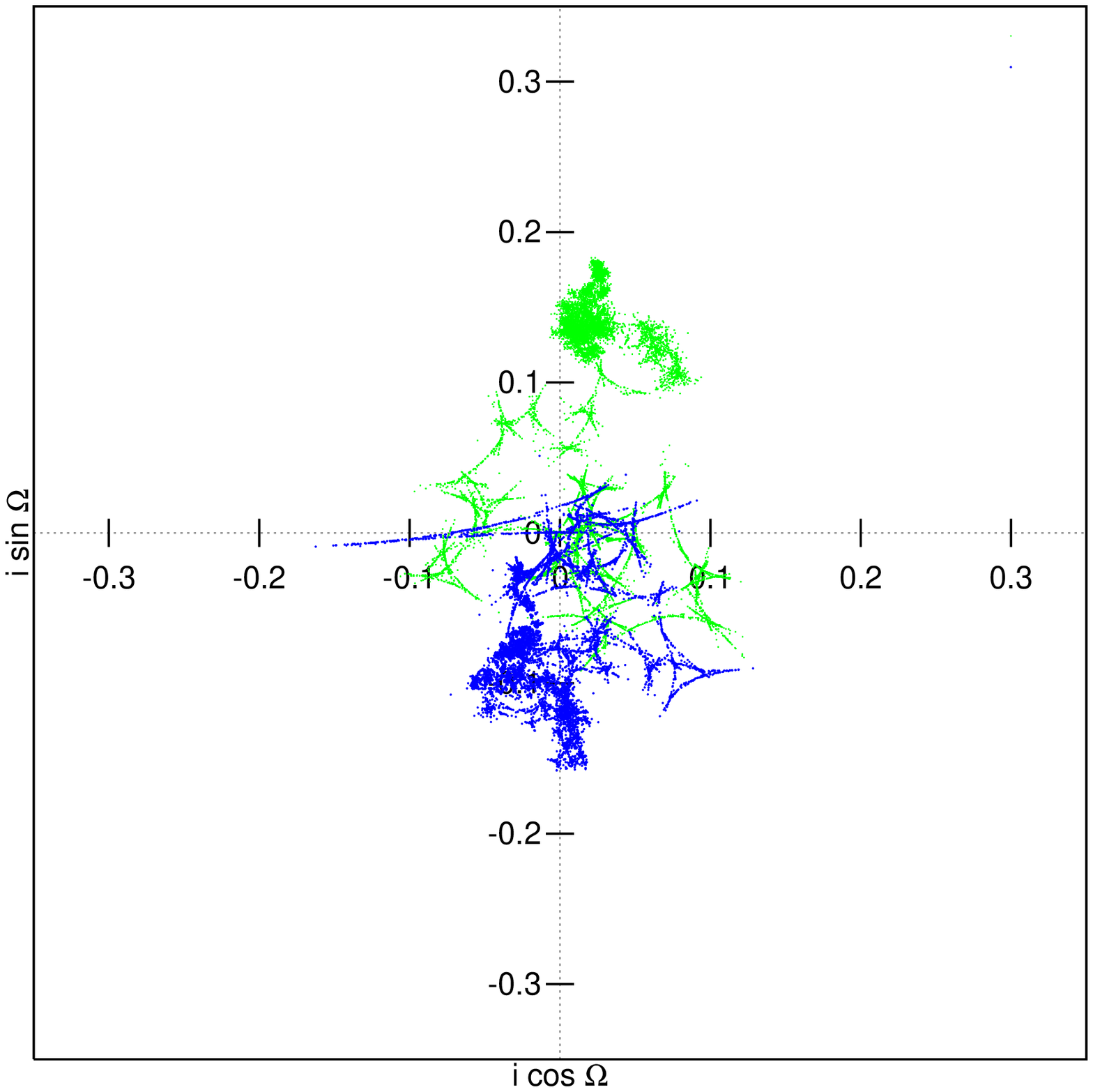}
\caption{
{\em Left panel: } Evolution of the inclination for models
${\cal P}$ (green) and ${\cal Q}$ (blue)
{\em Right panel: } Positions of the angular momentum vector in a polar representation
for the same models
}\label{incpol}
\end{figure*}
The direction of the orbital angular momentum vector of the binary
is specified by the quantities

\begin{equation}
\cos i=\textbf{l}\cdot \hat{\textbf{e}}_{z}/l\qquad 0\leq i<\pi
\end{equation}

\noindent
and

\begin{equation}
\left\{
\begin{array}{cc}
\cos\Omega=\textbf{K}\cdot \hat{\textbf{e}}_{x}/K & \textbf{K}\neq 0 \\
\Omega=0 & \textbf{K}=0 \\
\end{array}
\right. \qquad 0\leq\Omega<2\pi
\end{equation}

\noindent
where $\hat{\textbf{e}}_{x}$ and $\hat{\textbf{e}}_{z}$ are the
unit vectors of the reference coordinate system and $\textbf{K}$
the vector in the direction of the ascending node,
$\textbf{K}=\hat{\textbf{e}}_{z}\times\textbf{l}$, which represents
the definition of the inclination $i$ and the longitude of the
ascending node $\Omega$ following classical celestial mechanics.

The left side of Fig. \ref{incpol} shows the evolution of the
inclination for $W_{0}=6$ and $\omega_{0}=0.6$ for different
initial velocities. The inclination, typically in all simulations
undergoes comparatively strong changes during the dynamical friction
dominated regime, and remains passably stable or slightly drifting
during the hardening stage. Nevertheless, the total changes of the
inclination angle considering the whole simulation are rather
small. Sometimes a monotonic increase of the inclination angle occurs during
the dynamical friction dominated stage, and there were also simulations
in which the inclination dropped before reaching stable behaviour. This
can be seen in Tab. \ref{inctab}, where the maximum inclination $i_{\rm max}$ 
and the mean value seen in the stable phase $\bar{i}$ of each simulation
are listed: $i_{\rm max}$ and $\bar{i}$ can differ considerably.

In the right side of Fig. \ref{incpol}, the direction of the
orbital angular momentum vector is illustrated using polar
representation $(i\cos\Omega,i\sin\Omega)$. The $(0,0)$ coordinate
corresponds to a rotation of the binary in the $xy$-reference
plane. The extensive ripples are the result of periodic motion
before the binary is bound or when the binding is 
weak. With progressing evolution, the system concentrates in
confined cloud-like areas. The measured changes of direction of
the orbital angular momentum vector are consistent in order of
magnitude with the simulations of \citet{mil2001}, with the exception
of the inclination maverick $W_{0}=6$, $\omega_{0}=0.0$,
$v_{0}=0.136v_{c}$.

\section{Counter rotation}
\label{sec.dynamics2}

Applying exactly the same initial model for the field stars of the
corresponding model, the initial velocities of the IMBHs
were set contrary the rotation of the stellar system for the models
${\cal O}$ and ${\cal N}$.

\begin{figure*}
\includegraphics[width=6cm,angle=270]{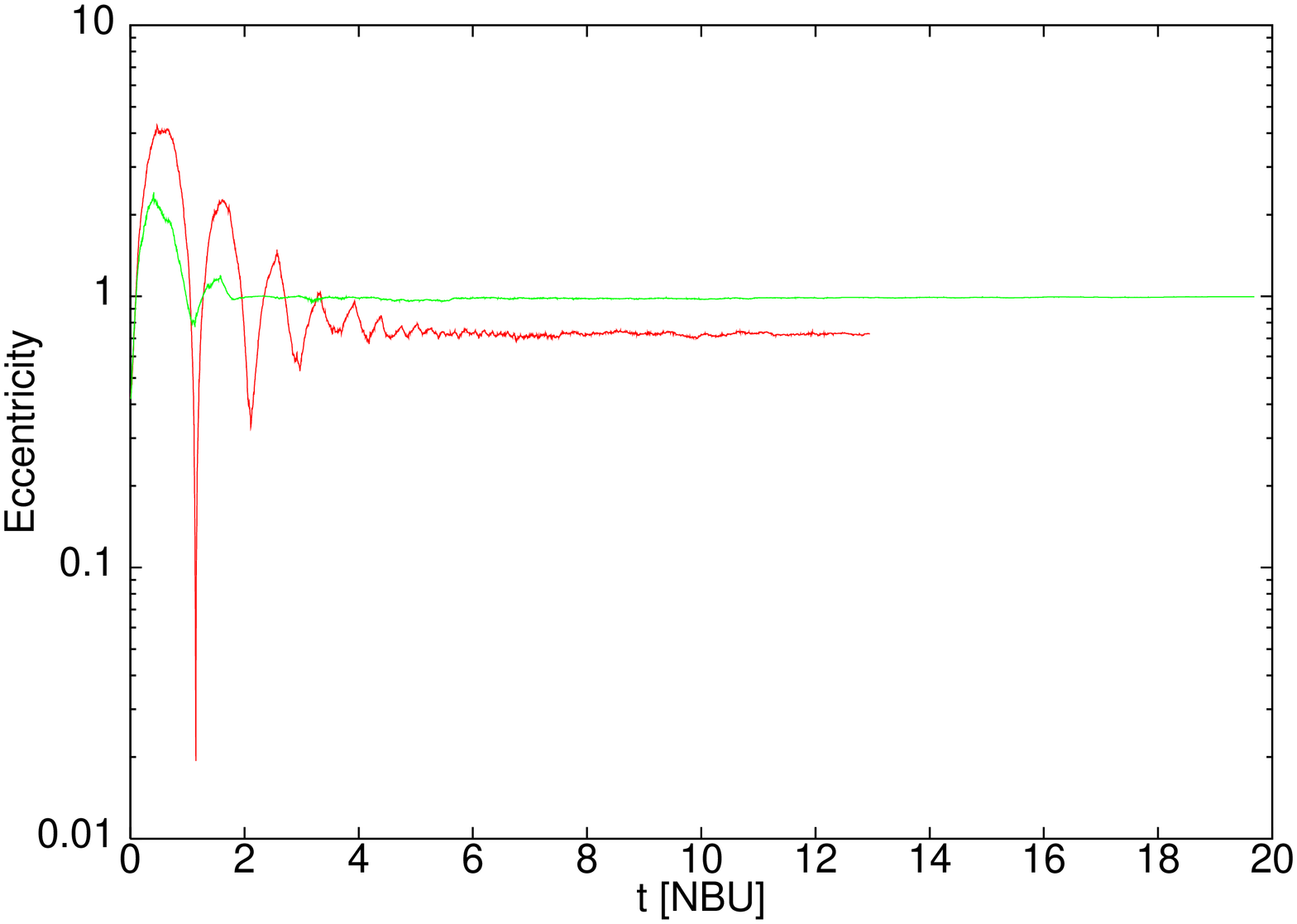}
\includegraphics[width=6cm,angle=270]{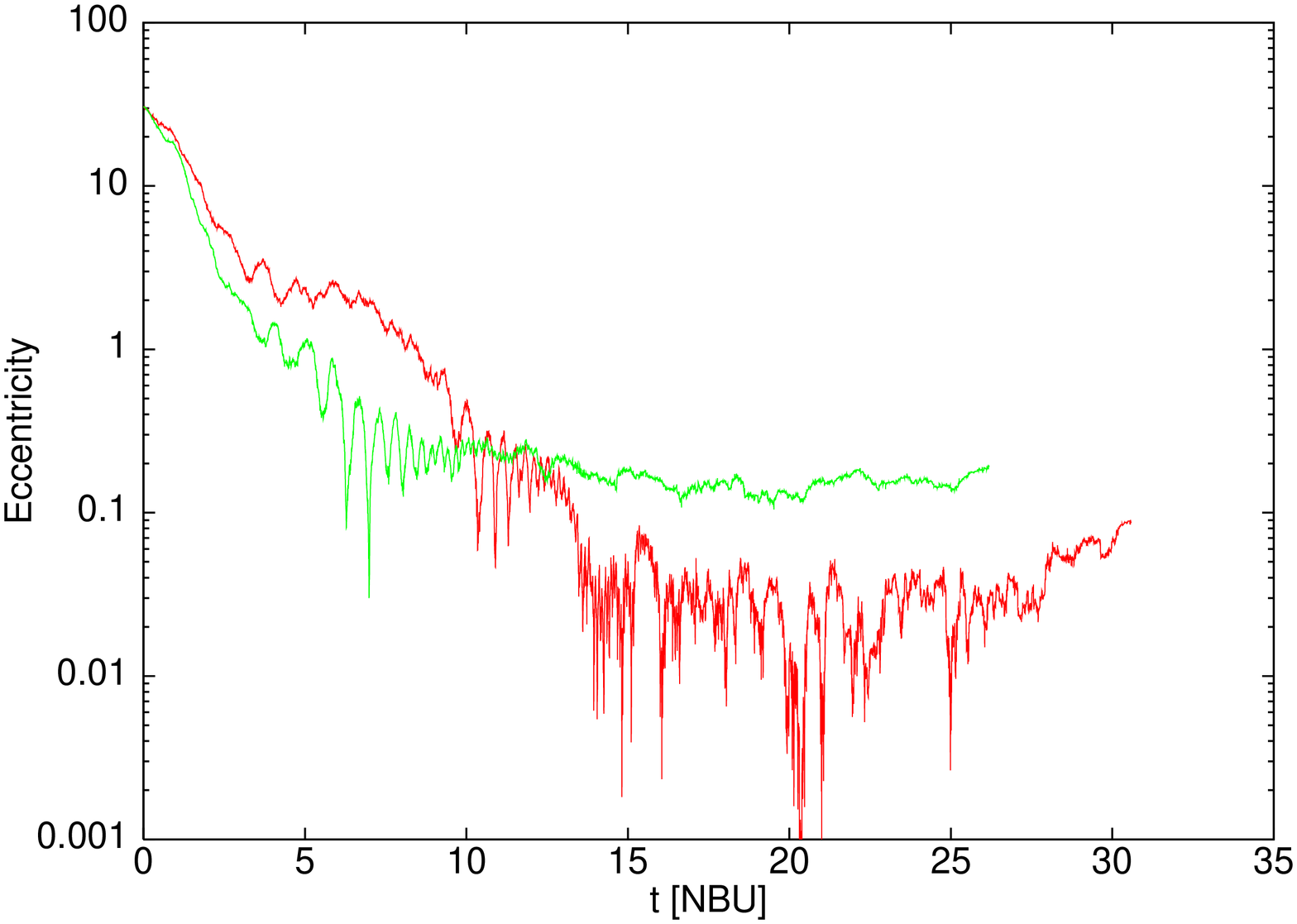}
\caption{
{\em Left panel:} 
Influence of rotation and counter-rotation on the
evolution of the eccentricity of the IMBHs. The upper curve (at later
times, green in the on-line version of the article) corresponds to 
model ${\cal O}$, for IMBHs initially set up
to be counter-rotating; the lower curve (red colour) is the same model with
co-rotation of the IMBHs.
{\em Right panel:} Same-same for model ${\cal N}$
}\label{counter}
\end{figure*}

The distinct influence of dynamical friction during the the first
time intervals is responsible for different results in eccentricity
evolution compared to the \emph{co}-rotating simulation (Fig.
\ref{counter}). In the case of $v_{0}=0.136v_{c}$, for
\emph{counter}-rotation an extremely high eccentricity
$\bar{e}=0,997$ is reached ($\bar{e}=0,728$ for
\emph{co}-rotating). In this scenario, the relative velocity
between a black hole and the field stars is increased at the
apoapsis as well as at the periapsis. Dynamical friction is very
efficient at the apoapsis of the individual black hole motion in
the unbound regime. Thus the black holes suffer a strong energy
loss and fall steeper to the centre than in the \emph{co}-rotating
simulation. The resulting high eccentric motion is kept into the
bound state. Applying an initial velocity $v_{0}=v_{c}$,  a
higher eccentricity $\bar{e}=0.160$ occurs compared to the
\emph{co}-rotating $\bar{e}=0.039$, but remains on a low level.

\section{Brownian motion}
The centre of mass (\rm CM) of a hardened binary is expected to to
perform an irregular motion in the central region of the stellar
system. This motion is often described by the concept of Brownian
motion, as it is characterised by a friction force (dynamical
friction) and a fluctuating force (as the result of scattering
events and encounters of field stars). Applying energy
equipartition in thermodynamic equilibrium, the mean square
velocity of the \rm CM of the binary $\langle v_{\rm CM}^{2}\rangle$ is
connected to the mean square velocity of the central field stars
$\langle v_{*}^{2}\rangle$ by
\begin{equation}\label{8}
\langle v_{\rm CM}^{2}\rangle=\frac{m_{\star}}{M_{\rm 2BH}}\langle
v_{*}^{2}\rangle
\end{equation}
where $M_{\rm 2BH}$ is the sum of the black hole masses.

However, the irregular motion of the \rm CM is to be distinguished
from the movement of a single massive particle since the binding
energy of the binary changes due to (super-elastic) scattering
events. The characteristics of the Brownian motion of a massive
black hole binary have been discussed in detail by
\citet{mer2001}, where $\langle v_{\rm CM}^{2}\rangle$ is expected
to be increased by a factor $\lesssim 2$, allowing for the
higher recoil velocities of a binary after super-elastic scattering
of field stars and for the decreased dynamical friction force on
the \rm CM, since the trajectories of the field stars are randomly
orientated in direction after such a process.

\begin{figure*}
\includegraphics[width=11cm,angle=270]{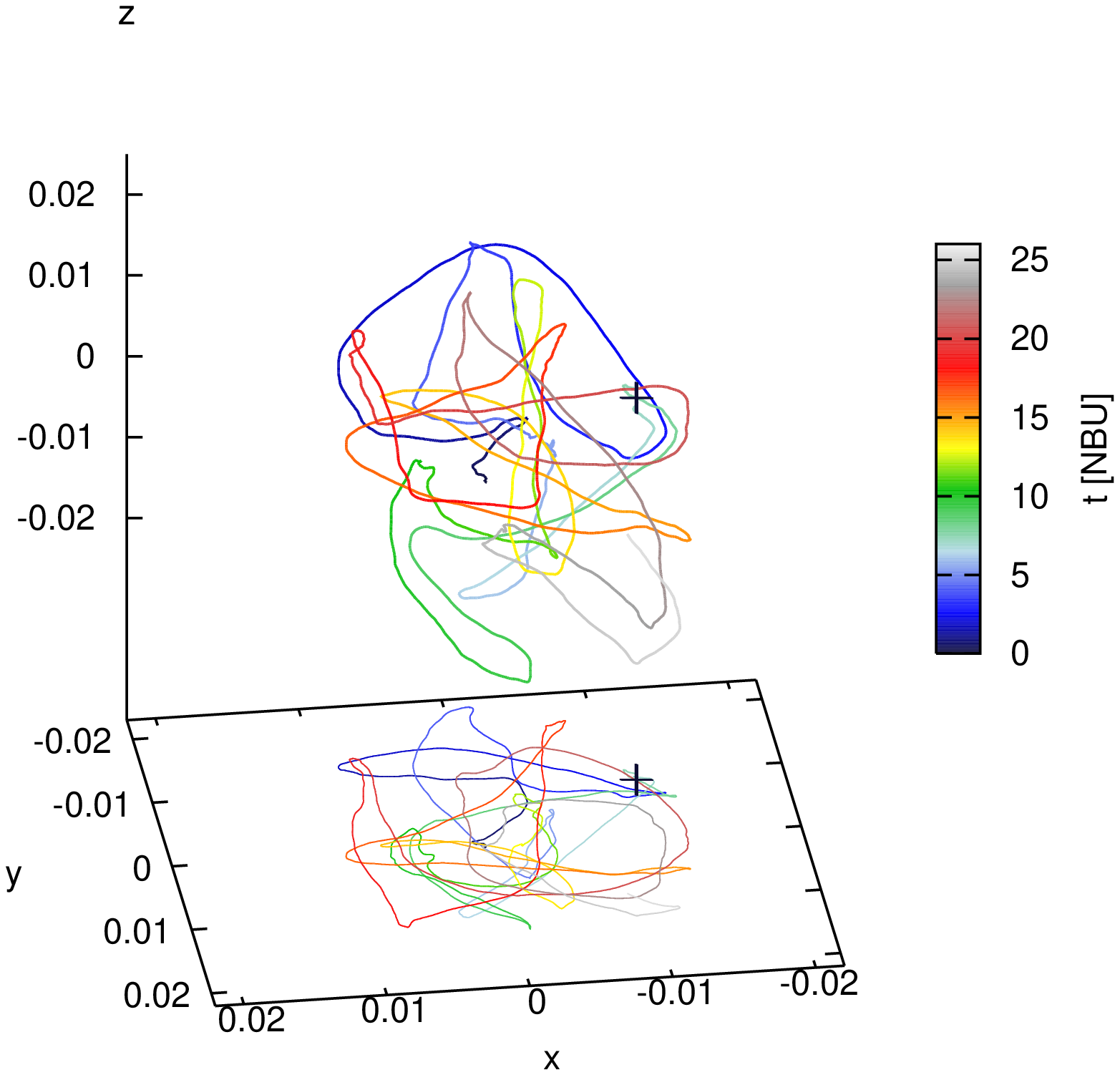}\\
\includegraphics[width=11cm,angle=270]{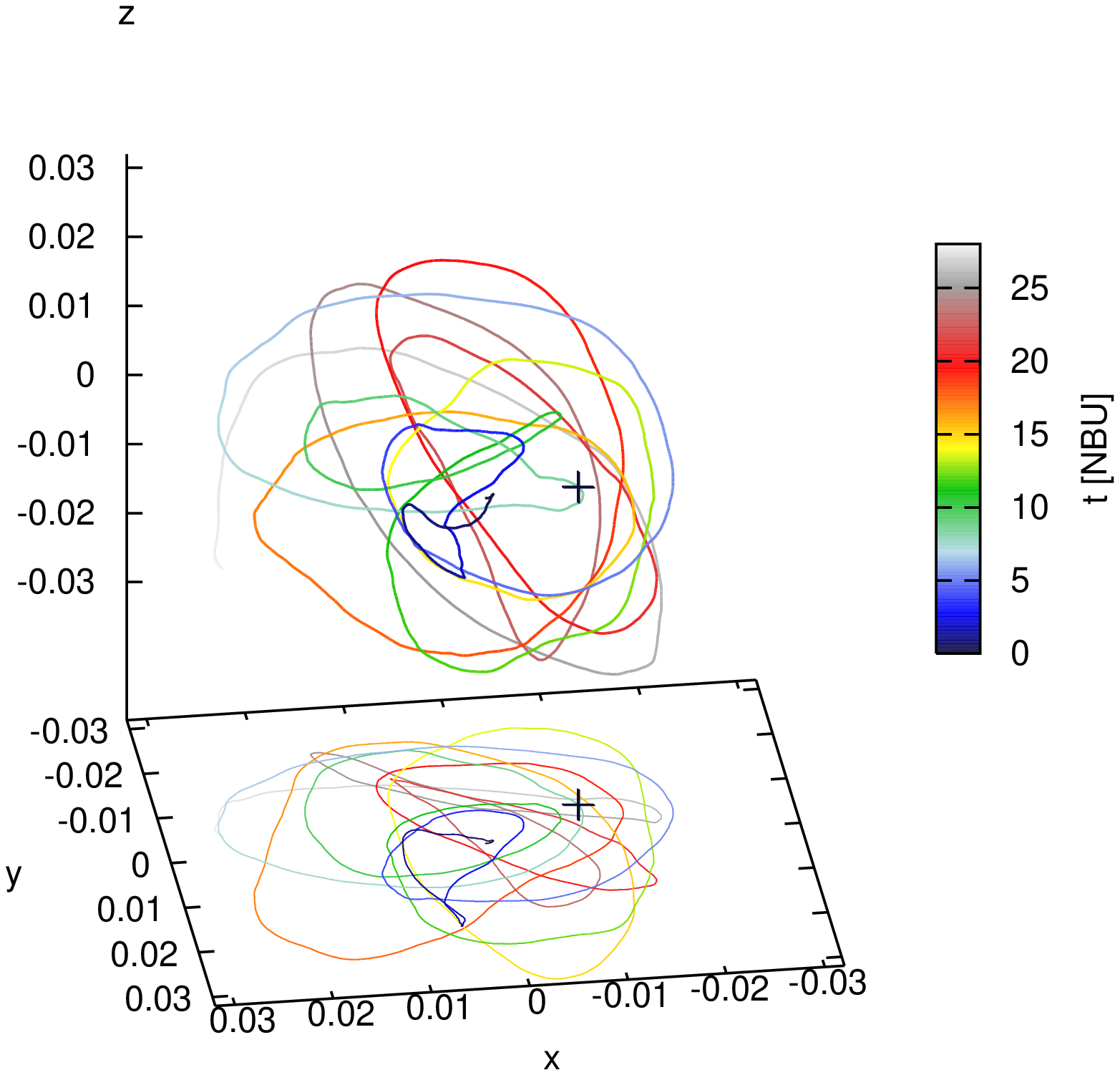}
\caption{{\em Upper panel:} Centre-of-mass trajectory of model ${\cal L}$
($\omega_{0}=0.0$). {\em Lower panel:} Same for model ${\cal Q}$ ($\omega_{0}=0.6$)
The cross indicates the moment in which there is a constant energy loss rate
($t \sim 7.6$ for $\omega_{0}=0.0$ and $t \sim 8.4$ for $\omega_{0}=0.6$)
}
\label{brmotion}
\end{figure*}

The CM motion was investigated for a series of King parameters.
Fig.\ref{brmotion} displays the CM movement during the whole simulation time
for King potential $W_0=6$, applying rotation parameters $\omega_0=0.0$ and
$\omega_0=0.6$, both with an initial IMBH velocity $v_0=0.136v_c$.  A color
gradient plot of the CM trajectory has been used to give an impression of the
temporal evolution. In the absence of rotation $\omega_0=0.0$, the
characteristics of an rather irregular motion appears, while with rotation
$\omega_0=0.6$, a turning motion of the CM occurs. This motion seems to follow
the rotation of the stellar system.

Elevated values of the CM mean velocity were found in
all simulations, quantified by the ratio $\langle
v_{\rm CM}^{2}\rangle/\langle v_{\rm equ}^{2}\rangle$ with $\langle
v_{\rm equ}^{2}\rangle =(m_{\star}/M_{\rm 2BH})\sigma_{0}^{2}$. The results of
three simulations are shown in Tab.\ref{browntab}. The postulated
factor $\lesssim 2$ is manifestly exceeded even without rotation,
but may be smaller if a a change of the central stellar velocity
dispersion $\sigma$ is taken into account, while calculations here
are based on the initial value $\sigma_{0}$ within the 1\%
Lagrangian radius of the stellar model. Nevertheless, a time
evolution of the mean square velocity indicated that a constant
value was not yet reached at the end of the simulations but may have
grown larger if runs had been continued.

\begin{table}
\centering
\begin{tabular}{c|c|c|c}
  \hline
  Model & $\langle v^{2}_{\rm CM}\rangle$
  & $\langle v_{\rm equ}^{2} \rangle$
  & $\langle v^{2}_{\rm CM}\rangle/\langle v_{\rm equ}^{2} \rangle$\\
  \hline
  ${\cal L}$ & $1.33\cdot 10^{-3}$ & $3.73\cdot 10^{-4}$ & 3.5 \\
  ${\cal N}$ & $2.79\cdot 10^{-3}$ & $3.66\cdot 10^{-4}$ & 7.6 \\
  ${\cal Q}$ & $2.18\cdot 10^{-3}$ & $3.32\cdot 10^{-4}$ & 6.6 \\
  \hline
\end{tabular}
\caption{Mean square velocities for the \rm CM of the binary $\langle
v^{2}_{\rm CM}\rangle$ compared to the equilibrium mean square
velocity $\langle v^{2}_{equ}\rangle$ of a single particle, for
some simulations. The $\langle v^{2}_{\rm CM}\rangle$ were calculated
over the blue epochs of section (\ref{brmotion}) and  over [11.60;30.58] in
the case $W_{0}=6$, $\omega_{0}=0.3$,
$v_{0}=v_{c}$.}\label{browntab}
\end{table}

\section{Gravitational Waves: Detectability of the systems with {\em LISA}}
\label{sec.GW}

\subsection{Post evolution of the binary of IMBHs}

\begin{figure*}
\resizebox{\hsize}{!}{\includegraphics[scale=1,clip]
{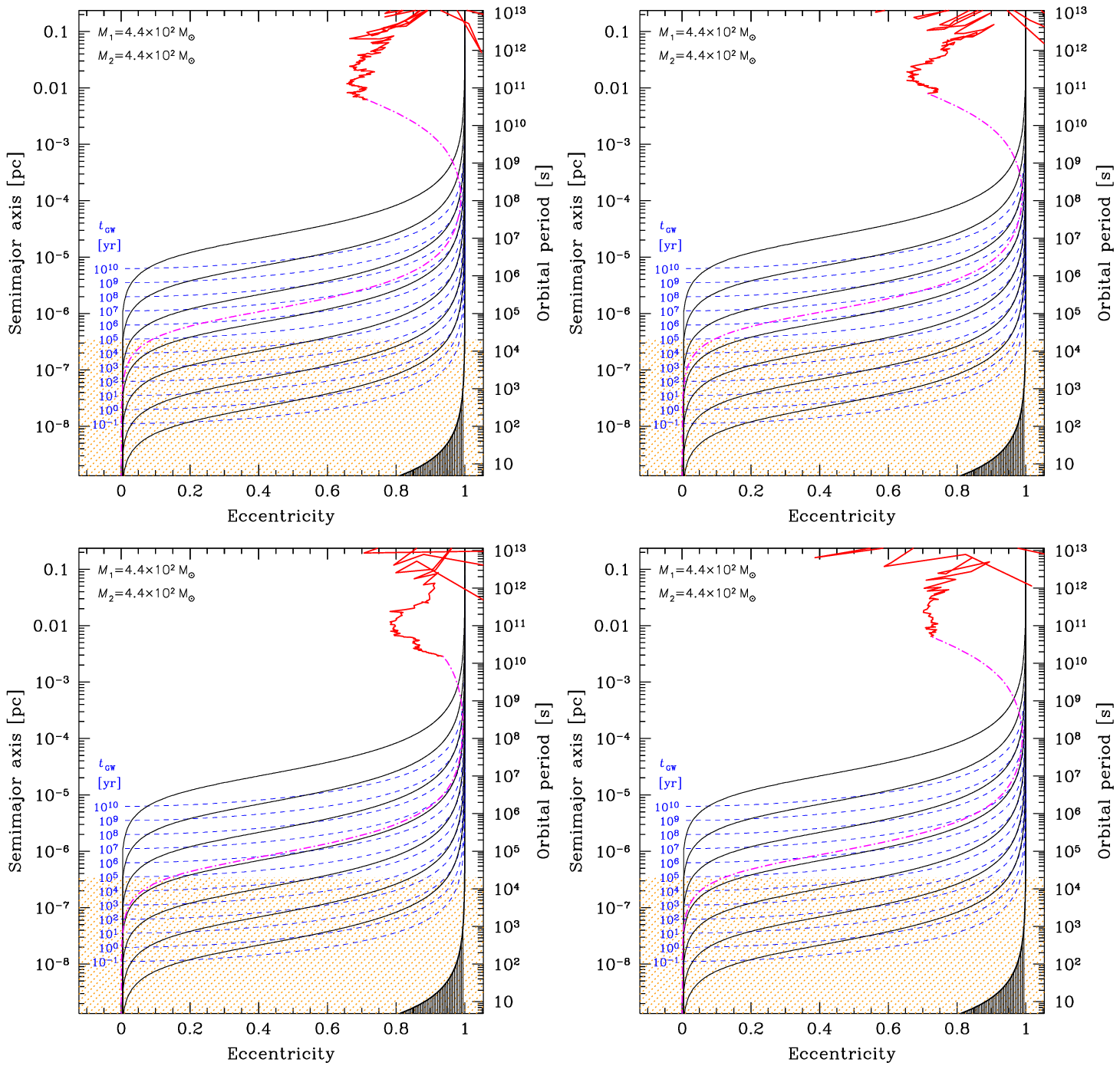}}
\caption{  
 Inspiral of the IMBH binary of Models~${\cal C, F, L}$ and ${\cal O}$ from the
 top to the bottom and from the left to the right. We show in this figure the
 evolution of the binary in the eccentricity--semi-major axis plane. The zigzag
 curves (red in the on-line version) depict the results of the $N-$body
 simulations. Initially the IMBHs do not constitute a bound system and
 therefore the eccentricity of the system is ill-defined, as explained at the
 beginning of Section (\ref{sec.EvolBindEnergy}). This is causing the
 initial values of $e$ become $> 1$. We then make a semi-analytical expansion
 of the evolution starting from the last point in the numerical simulations
 (dot-dashed curve, magenta in the on-line version of the paper)
 by taking into accound the Physical propierties of the system at that moment:
 semi-major axis, eccentricity, velocity dispersion of the stellar system and
 stellar density, as shown in Table \ref{tab.EndPhysPropSystem}. Thus, we evolve the
 binaries until they enter the LISA bandwidth, which we depict as a
 lightly shaded area (orange in the on-line version). The black solid curves
 correspond to the trajectories due only to the emission of GWs \citep{pet1964}
 and we additionally show the corresponding inspiral $t_{\rm GW}$ for $10^{10}$
 yrs, $10^9$ yrs etc. As shown in \citep{ASF06},
 one recovers partially the 
 $N-$body results with the semi-analytical approach if one starts at a previous 
 point in the curve corresponding to the numerical simulation. The black-shaded
 region on the right corresponds to the last stable circular orbit.
\label{fig.ModelsCFLO}
}
\end{figure*}

The advantage of direct-summation codes, accuracy, is at the price of
performance.  We have chosen $N-$body in order to investigate this problem but
in order to analyse the ulterior evolution of the binary down to a GW frequency
observable by LISA, we have to resort to alternative schemes. If we were to
integrate the binary system until the orbital period of the binary is within
the range of observations of LISA, we would have to leave the simulations
running for months. This is not desirable for obvious reasons.  Instead, we
recur to a semi-analytical method to evolve the orbital elements of the binary taking
into account the dynamics and the GW emission of the system, as introduced in
\cite{ASF06}: we stop the direct-summation calculation after the initial strong
fluctuating phase; when the eccentricity achieves a {\em steady} value.  In order to
locate in the LISA bandwidth the position of the binary, we employ the results
of the direct-summation simulation and extend them with a semi-analytical
method. The dynamics will, in general, tend to increase the eccentricity,
whilst the emission of GWs circularies the orbit. These two processes are
competitive. The basic idea to further evolve the binary is to split the
evolution of both the semi-major axis and the eccentricity in two
contributions, one driven by the dynamical interactions with stars (subscript
{\em dyn}) and another by the emission of GWs (subscript {\em GW}),

\begin{equation}
\frac{da}{dt}=\left(\frac{da}{dt}\right)_{\rm dyn}+\left(\frac{da}{dt}\right)_{\rm GW},~
\frac{de}{dt}=\left(\frac{de}{dt}\right)_{\rm dyn}+\left(\frac{de}{dt}\right)_{\rm GW}
\label{eq.split}
\end{equation}

\noindent
The {\em GW} terms are as given by the approach used in \citet{pet1964}. 
As for the dynamical part, we resort to the scheme described in \cite{Quinlan96}.
For more details about this approach, see Section 3 of \cite{ASF06}.

\begin{table*}
\begin{tabular}{c|c|c|c|c|c|c|c|c|c|c|c}
\hline Simulation & Time & $R_c$ & $N$ & $\sigma^2_R$ & $\sigma^2_T$ & $\sigma_{tot}$ & $n$ & $\bar{m}$ & $\rho$ & $\rho*$ & $\sigma_{tot}$* \\
\hline
$\cal C$ & 36.0 & 0.376 & 7778 & 0.255 & 0.304 & 0.929 & 34878.9 & $1.82*10^{-5}$ & 0.635 & 0.544 & 0.860\\
$\cal F$ & 33.0 & 0.376 & 7627 & 0.253 & 0.261 & 0.880 & 34201.8 & $1.82*10^{-5}$ & 0.622 & 0.534 & 0.943\\
$\cal L$ & 25.0 & 0.216 & 4209 & 0.476 & 0.782 & 1.428 & 99739.3 & $2.03*10^{-5}$ & 2.025 & 1.555 & 1.575\\
$\cal O$ & 12.0 & 0.223 & 4270 & 0.383 & 0.265 & 0.956 & 91828.0 & $2.03*10^{-5}$ & 1.864 & 1.433 & 0.967\\
\hline
\end{tabular}
\caption{All values are given in $N-$body units. In the table $R_c$ is the core radius,
$N$ the number of stars within the core radius, including the two IMBHs, 
$\sigma^2_R$ the radial velocity dispersion (squared),
$\sigma^2_T$ the tangetial velocity dispersion (squared, 1D),
$\sigma_{tot}=\sqrt{\sigma^2_R+2\sigma^2_T}$ the velocity dispersion (3D, including IMBHs),
$n$ the particle density within the core radius (including IMBHs),
$\bar{m}$ the average particle mass within the core radius (including IMBHs),
$\rho$ the average mass density within the core radius (including IMBHs),
$\rho*$ the average mass density within the core radius (excluding IMBHs),
$\sigma_{tot}*$ the velocity dispersion within the core radius (3D, excluding 2BH)
\label{tab.EndPhysPropSystem}
}
\end{table*}

In Fig.(\ref{fig.ModelsCFLO}) we depict the results of this approach for the
models ${\cal C, F, L}$ and ${\cal O}$. We have chosen them for being the most
interesting ones from a dynamical point of view, since they have the largest
eccentricities by the moment of entrance in the observatory's sensitivity
window. Also, from the standpoint of detectability, these are the most
challenging ones due to the same reason. We can see that in all four models the
binary of IMBHs enters the bandwith with a residual eccentricity which, even if
rather mild, is not negligible, from the point of view of detection.

\subsection{Event rates}

\cite{FregeauEtAl06} estimated the detection of binary IMBHs by LISA for the
single cluster channel. They assumed that any cluster undergoing a collisional
runaway, such as those found in the Monte Carlo numerical simulations of
\citet{GFR06}, form a two very massive stars. These evolve separately and
eventually may collapse and build two IMBHs separately, so that the IMBHs do
{\em not} coalesce in the process of their formation, but are born
independently.

\cite{ASF06} calculated the event rate for formation of binaries of IMBHs based
on the results of \cite{FregeauEtAl06} in the scenario of two colliding
clusters, the double cluster channel (see their Section 4, also for a more
detailed explanation of the following events). Their work assumed that the
IMBHs were already present at the centres of the two clusters undergoing the
crash. When the two clusters merge, the IMBHs are drawn to the centre due to
dynamical friction and constitute a binary which eventually coalesce. The
difference in the calculation of event rates of \cite{ASF06} and
\cite{FregeauEtAl06} is the number of IMBHs formed per cluster and the
requirement for the host clusters to merge.

In both estimations, the probability that a cluster evolves to the runaway
phase was set to 0.1 as an illustrative case, though it can be as high as 0.5
\citep{FreitagRB06}. Both works assumed that a runaway always leads to the
formation of an IMBH. We refer the reader to Section 4 of \cite{ASF06} and
\cite{FregeauEtAl06} for a detailed explanation and exposition of the
uncertainties in the calculations.  To summarise, the \cite{FregeauEtAl06}
results for the LISA detector are

\begin{align}
\Gamma_{\rm Freg}|_{\rm opt} & \in [200,\,250]\,{\rm yr}^{-1} \\\nonumber
\Gamma_{\rm Freg}|_{\rm pes} & \in [40,\, 50]\,{\rm yr}^{-1}.
\end{align}

\noindent
Where the subscript ``Freg'' stands for \cite{FregeauEtAl06}, the subscript``opt'' 
for the optimistic estimation, assuming that
the probability for a cluster evolving to the runaway stage is 0.5, and the
subscript ``pes'' stands for pesimistic, which is the result of using 0.1 instead.
\cite{ASF06} find the following results, where we use a nomenclature similar as above
and set the probability for the host clusters to merge to 1 (these would decrease by
a factor 10 if one was to use 0.1 instead, see discussion about
the ``UCDG channel'' in their work)

\begin{align}
\Gamma_{\rm ASF}|_{\rm opt} & \in [100,\, 125]\,{\rm yr}^{-1} \\\nonumber
\Gamma_{\rm ASF}|_{\rm pes} & \in [4,\, 5]\,{\rm yr}^{-1}.
\end{align}

\noindent
So that the contribution to the total number of binaries of IMBHs from both channels
is

\begin{align}
\Gamma_{\rm tot}|_{\rm opt}  & \in [300,\,375]\,{\rm yr}^{-1} \\\nonumber
\Gamma_{\rm tot}|_{\rm pes}  & \in [44,\, 55]\,{\rm yr}^{-1}
\end{align}

\noindent
Or, in the unlikely very pesimistic situation of having the host clusters to merge with a 0.1 probability,
we have the following ``optimistic--pesimistic'' and ``pesimistic--pesimistic'' results

\begin{align}
\Gamma_{\rm tot}|_{\rm opt}^{\rm pes}  & \in [210,\,262.5]\,{\rm yr}^{-1} \\\nonumber
\Gamma_{\rm tot}|_{\rm pes}^{\rm pes}  & \in [40.4,\, 50.5]\,{\rm yr}^{-1}
\end{align}

These results are encouraging enough that we address the parameter estimation of the
sources. We describe the methods and results in the next sections.

\subsection{The gravitational waveform}

To investigate the detectability of these sources for LISA, we use the
restricted post-Newtonian (PN) approximation for the GW, where we assume 2-PN
corrections to both the conservative and adiabatic dynamics of the system, but
conserve the amplitude at the dominant Newtonian order.  With this in mind, the
waveform polarisations for non-spinning eccentric binaries at the detector are
given by (in units of $G=c=1$)~(\cite{dgi2004})

\begin{eqnarray}
h_+(t) &= &\frac{M\eta}{ D_L}  \left \{\left(1+\cos ^2\iota \right) \bigg[\cos 2 \varphi
    \left(-\dot{r}^2+r^2 \dot{\varphi
   }^2+\frac{M}{r}\right) \right.  \nonumber\\
&  +&2 r\, \dot{r}\, \dot{\varphi } \sin 2 \varphi \bigg ] 
\left. +\left(-\dot{r}^2-r^2 \dot{\varphi }^2+\frac{M}{r}\right) \sin ^2 \iota \right \} \\
h_\times(t) &=& -\frac{2 M \eta}{ D_L}  \cos \iota  \bigg\{\left(-\dot{r}^2+r^2
   \dot{\varphi }^2+\frac{M}{r}\right) \sin 2 \varphi \nonumber \\
&  -&2 r\,\dot{r}\, \dot{\varphi } \cos 2 \varphi \bigg\} \, ,
\end{eqnarray}

\noindent
where $\cos\iota=\bf\hat{L}\cdot\hat{n}$.  Here, ${\bf
\hat{L}}$ is the direction of the binary's orbital angular momentum
and ${\bf\hat{n}}$ is the direction from the observer to the source
(such that the GWs propagate in the ${-\bf\hat{n}}$ direction). $M$ is the
total mass of the system, $\eta = M_1 M_2 / M^2$ is the symmetric mass ratio and
$D_L$ is the luminosity distance to the source .  The components $(r, \varphi)$
denote the orbital radius and phase of the system (also referred to as the true
anomaly), and are schematically described by the equations~(\cite{hind2008})

\begin{eqnarray}
r/M &=& r_\mathrm{0PN} x^{-1} + r_\mathrm{1PN} + r_\mathrm{2PN} x + \mathcal{O}(x^2) \\
M \dot \phi &= &  \dot \phi_\mathrm{0PN}x^{3/2} + \dot \phi_\mathrm{1PN} x^{5/2} + \dot \phi_\mathrm{2PN} x^{7/2}+
 \mathcal{O}(x^{9/2}) \label{eqn:phidotpn} \\
l &=&   l_\mathrm{2PN} + l_\mathrm{2PN} x^2   + \mathcal{O}(x^3) \label{eqn:keplerpn} \\
M \dot l &=& Mn = x^{3/2} + \dot l_\mathrm{1PN}x^{5/2} + \dot l_\mathrm{2PN} x^{7/2}  + \mathcal{O}(x^{9/2}) \label{eqn:ldotpn} \, ,
\end{eqnarray}
where $l$ is the mean anomaly and $n=2\pi/P$ is the mean motion, where $P$ is the orbital period defined as the time to go from pericenter to pericenter.  Due to precession effects, this is different from the time taken to go from $\varphi$ to $\varphi+2\pi$.  The dot represents the time derivative, e.g. $\dot\varphi=d\varphi/dt$.  We also define the invariant PN coefficient $x =(M\omega)^{2/3}$, where the angular frequency is defined as $\omega =  (2\pi+\Delta\varphi)/P$ and $\Delta\varphi$ is the precession angle of the pericenter per period.  We should note that the coefficients in the equations presented above and below are in general functions of the instantaneous eccentricity $e$ and the eccentric anomaly $u$. In the limit of the eccentricity $e\rightarrow 0$, we reclaim the circular orbit case where $\omega=\dot\varphi$. The adiabatic evolution of $x$ and $e$ are given by
\begin{eqnarray}
M \dot x = \dot x_\mathrm{0PN} x^5 + \dot x_\mathrm{1PN} x^{6} 
  + \dot x_\mathrm{2PN} x^{7} + \mathcal{O}(x^{15/2}) \\
M \dot e = \dot e_\mathrm{0PN}  x^4 + \dot e_\mathrm{1PN} x^{5} +  \dot e_\mathrm{2PN} x^{6} + \mathcal{O}(x^{13/2}) \, .
\end{eqnarray}

\noindent
where we point the reader to the appendix of ~\cite{hind2008} for the complete
description of the above equations.  The waveforms are generated as follows :
after we have evolved the above equations for $x$ and $e$, we numerically
integrate Equation~\ref{eqn:ldotpn} to obtain $l(t)$.  We then use this value
to solve the post-Newtonian Kepler's Equation~\ref{eqn:keplerpn}, which is a
transcendental equation in $u(t)$.  Once we have $u(t),e(t)$ and $x(t)$, we can
then calculate $r(t), \dot{r}(t), \varphi(t), \dot{\varphi}(t)$, where we
calculate the integral of $\dot\varphi(t)$ and the derivative $\dot r(t)$
numerically.

In this work, to fully describe the GW polarisations we use the following
parameter set : $\vec\lambda=\left\{\ln(M_c), \ln(\mu), \ln(D_L), \ln(a_0),
\cos\theta, \phi, e_0, \cos\iota, \psi \right\}$, where $M_c=M\eta^{3/5}$ is
the chirp-mass, $\mu=M\eta$ is the reduced mass, $D_L$ is the luminosity
distance, $a_0$ is the initial semi-major axis of the orbit, $(\theta,\phi)$
are the co-latitude and longitudinal position of the source in the sky, $e_0$
is the initial eccentricity, $\iota$ is the inclination of the orbital plane
and $\psi$ is the GW polarisation angle.

\subsection{Detector response and parameter error estimation}

LISA can be thought of as pair of co-located detectors rotated with respect to
each other by an angle of 45 degrees.  In the Low Frequency Approximation
(LFA) \citep{cutler98}, we can write the individual detector responses as

\begin{equation}
h_i(t) = h_{+}(\xi(t))F^{+}_i(t)+h_{\times}(\xi(t))F^{\times}_i(t), 
\end{equation}
where $\xi(t)$  is the phase shifted time parameter defined by
\begin{equation}
\xi(t) = t - R_{\oplus}\sin\theta\cos\left(\alpha(t) - \phi\right).
\end{equation}
Here $R_{\oplus}$ corresponds to one AU and $\alpha(t) = 2\pi f_m t +\kappa$, where the LISA modulation frequency is $f_m = 1/year$ and $\kappa$ is the initial ecliptic longitude of the guiding center of the LISA constellation.  The functions $F^{+,\times}(t;\theta, \phi, \psi)$ are the beam pattern functions of the detector~(\cite{cr2003}).

Given sources $h(t)$ and $g(t)$ we can define a noise weighted inner product
\begin{equation}\label{eqn:scalarprod}
\left<h\left|g\right.\right> =2\int_{0}^{\infty}\frac{df}{S_{n}(f)}\, \tilde{h}(f)\tilde{g}^{*}(f) + \tilde{h}^{*}(f)\tilde{g}(f).
\end{equation}
where $\tilde{h}(f)$ is the Fourier transform of the signal, an asterisk denotes a complex conjugate term and $S_n(f)$ is the one-sided noise spectral density of the detector.  For this study, we use a noise curve given by
\begin{equation}
S_{n}(f) = S_{n}^{\rm instr}(f) + S_{n}^{\rm conf}(f)
\end{equation}
where the instrumental noise $S_n(f)$ is given by
\begin{eqnarray}
S_{n}^{\rm instr}(f) &= &\frac{1}{4L^{2}}\left[ 2S_{n}^{\rm pos}(f) \left(2+\left(\frac{f}{f_{*}}\right)\right) + 8 S_{n}^{\rm acc}(f) \right. \nonumber\\ \\
&\times & \left.   \left(1+\cos^2\left(\frac{f}{f_{*}}\right)\right)  \left( \frac{1}{(2\pi f)^{4}} + \frac{(2\pi 10^{-4})^{2}}{(2\pi f)^{6}}   \right)   \right] . \nonumber
\end{eqnarray}
\citep{cr2003}.
Here $L=5\times10^{6}$ km is the arm-length for LISA,  $S_{n}^{\rm pos}(f) = 4\times10^{-22}\,m^{2}/Hz$ and $S_{n}^{\rm acc}(f) = 9\times10^{-30}\,m^{2}/s^{4}/Hz$ are the position and acceleration noises respectively.  The quantity $f_{*}=1/(2\pi L)$ is the mean transfer frequency for the LISA arm.   Note that we have also included a reddened noise term which steepens the noise curve between $10^{-4}$ and $10^{-5}$ Hz.  To model the galactic or confusion noise we use the following confusion noise estimate derived from a Nelemans, Yungelson, Zwart (NYZ) galactic foreground model \citep{NYZ, TRC}
\begin{equation}
S_{n}^{\rm conf}(f) = \left\{ \begin{array}{ll} 10^{-44.62}f^{-2.3} & 10^{-4} < f\leq 10^{-3} \\ \\ 10^{-50.92}f^{-4.4} & 10^{-3} < f\leq 10^{-2.7}\\ \\ 10^{-62.8}f^{-8.8} &  10^{-2.7} < f\leq 10^{-2.4}\\ \\ 10^{-89.68}f^{-20} &  10^{-2.4} < f\leq 10^{-2}  \end{array}\right.,
\end{equation}
where the confusion noise has units of $Hz^{-1}$. 

Using the noise weighted inner product, we can define the optimal signal to noise  ratio (SNR) by
\begin{equation}
\left<h\left|h\right.\right>_{\rm opt} =4\int_{0}^{\infty}\frac{df}{S_{n}(f)}\,| \tilde{h}^{2}(f)|.
\end{equation}
We can also define the Fisher information matrix (FIM) by
\begin{equation}
\Gamma_{\alpha\beta} = \left<\partial_{\alpha}h\left|\partial_{\beta}h\right.\right>.
\end{equation}
where the theoretical standard deviation error estimate in parameter recovery is given as
\begin{equation}
\sigma_{\alpha} = \sqrt{\left(\Gamma_{\alpha\alpha}\right)^{-1}}
\end{equation}
The derivatives of the waveforms appearing in the FIM are generated numerically.  We refer the reader to \citep{PC2008} for the intricacies in the numerically calculation of the FIM.

\begin{table*}
\centering
\begin{tabular}{c|c|c|c|c|c|c|c}
  \hline
  Model & $c_0$ & $c_1$ &$c_2$
  & $a_{0}^{\rm min}\times 10^{-8}pc$ & $a_0^{\rm max}\times10^{-8}pc$ & $e_{0}^{\rm min}$ &$e_{0}^{\rm max}$ \\
  \hline
 ${\cal C}$  & $1.1366\times10^{-3}$ & $9.6414\times10^{4}$ & $9.4069\times10^{11}$ & 3.6123 & 10.669 & $3.5924\times10^{-3}$ & $1.9942\times10^{-2}$ \\
 ${\cal F}$  & $6.9767\times10^{-4}$ & $6.3737\times10^{4}$ & $6.9846\times10^{11}$ & 3.6107 & 9.8673 & $2.5227\times10^{-3}$ & $1.2389\times10^{-2}$ \\
 ${\cal L}$  & $1.1591\times10^{-3}$ & $1.0606\times10^{5}$ & $1.1698\times10^{12}$ & 3.7049 & 9.3739 & $4.3852\times10^{-3}$ & $1.9056\times10^{-2}$ \\
 ${\cal O}$  & $8.6073\times10^{-4}$ & $8.4893\times10^{4}$ & $1.0181\times10^{12}$ & 3.8524 & 9.0204 & $3.9227\times10^{-3}$ & $1.5081\times10^{-2}$ \\
 \hline
 \end{tabular}
\caption{Quadratic power law coefficients for the four models ${\cal C,F,L,O}$, as well as minimum and maximum ranges for the initial semi-major axes $a_0$ and eccentricity $e_0$.}\label{tab:ecca}
\end{table*}

\subsection{Sampling the parameter space}

Figure 10 provides the semi-major axes, eccentricities and orbital periods for
the models ${\cal C, F, L}$ and ${\cal O}$, assuming an equal mass IMBH binary
with individual rest-frame masses of $M_i = 440\,M_{\odot}$.  To sample the
parameter space we ran a 1000 iteration Monte Carlo simulation over the
parameters $\left\{a_0, e_0, \theta, \phi, \iota, \psi\right\}$, while keeping
the luminosity distance and redshifted mass parameters constant.  For the
angular parameters we assume $(\cos\theta, \cos\iota) \in[-1, 1]$ and $(\phi,
\psi) \in [0, 2\pi]$ and we then draw uniformly from these ranges.  

As the IMBHs in the study are quite low mass as compared to SMBHs, we need to
ensure that the sources are detectable. Given the masses of the systems in
question, the GW frequency at the last stable circular orbit is between 4-6 Hz,
which is well outside the frequency range of LISA.  Therefore, we placed all
sources at a common distance of $D_L$=100 Mpc and required a SNR greater than
5.  At this distance, the parameter values observed by LISA are the redshifted
rather than rest frame values.  To account for this, the measured total mass is
$M(z)=(1+z)M$ and the measured GW frequency of the waveform is $f_{\rm gw}(z)=f_{\rm gw} / (1+z)$.    
In
this study we use the following relation
between redshift, $z$, and luminosity distance, $D_{L}$ :

\begin{eqnarray}
D_{L} = \frac{c(1+z)}{H_{0}}\int_{0}^{z}\,dz'\left[\Omega_{R}
\left(1+z'\right)^{4} +
 \Omega_{M} \left(1+z'\right)^{3}+\Omega_{\Lambda} \right]^{-1/2},
\end{eqnarray}

\noindent
where we assume WMAP values of values of $(\Omega_{R}, \Omega_{M},
\Omega_{\Lambda}) = (4.9\times10^{-5}, 0.27, 0.73)$ and a Hubble's
constant of $H_{0}$=71 km/s/Mpc.

We also decided to enforce a maximum possible GW
frequency of 3 mHz to ensure the fidelity of the LFA \citep{cr2003,kc2009}.
Thus, using the information in Figure 11, we evolved the sources from first
entering the LISA bandwidth to the point where they achieved the required SNR
threshold.  For this we assumed a 3 year mission lifetime for LISA.  
 
Using the above constraints it was possible to find minimum and maximum values
of ($a_0, e_0$) which satisfied both the SNR and maximum frequency
constraints.  Furthermore to sample this parameter sub-space we found it was
possible to relate the eccentricity and semi-major axis for the four models
according to a quadratic law, i.e.  \begin{equation} e = -c_0 + c_1\,a_0 +
c_2\,a_{0}^{2}, \end{equation} where we provide the coefficients $c_i$, the
maximum and minimum values of both $a_0$ and $e_0$ in Table~\ref{tab:ecca}.
Thus, by uniformly sampling $a_0$, we also have a corresponding sample in
$e_0$.  We can see from Table~\ref{tab:ecca} that although the binaries have
appreciable eccentricities when they first enter the LISA band, i.e $e\sim
0.1-0.15$, by time the systems become observable the eccentricity has dropped
to $e\sim 0.012-0.019$.

While the Monte Carlo is carried out using the sky coordinates of the source, a
more interesting quantity to quote is LISA's angular resolution for each
source.  We define the angular resolution as

\begin{equation}
\Delta\Omega = 2\pi\sqrt{\Sigma^{\theta\theta}\Sigma^{\phi\phi}-
                \left(\Sigma^{\theta\phi}\right)^{2}},
\end{equation}

\noindent
where

\begin{equation}
\Sigma^{\alpha\beta} = \left<\Delta\lambda^{\alpha}\Delta\lambda^{\beta}\right> = 
                       \left(\Gamma_{\alpha\beta}\right)^{-1}.
\end{equation}

\noindent
We can now define the quantities appearing in the angular resolution as
$\Sigma^{\theta\theta} = \left<\Delta\cos\theta\Delta\cos\theta\right>$,
$\Sigma^{\phi\phi} = \left<\Delta\phi\Delta\phi\right>$ and
$\Sigma^{\theta\phi} = \left<\Delta\cos\theta\Delta\phi\right>$.  The angular
resolution has units of steradians.

\begin{figure*}
\resizebox{\hsize}{!}{\includegraphics[scale=1,clip]
{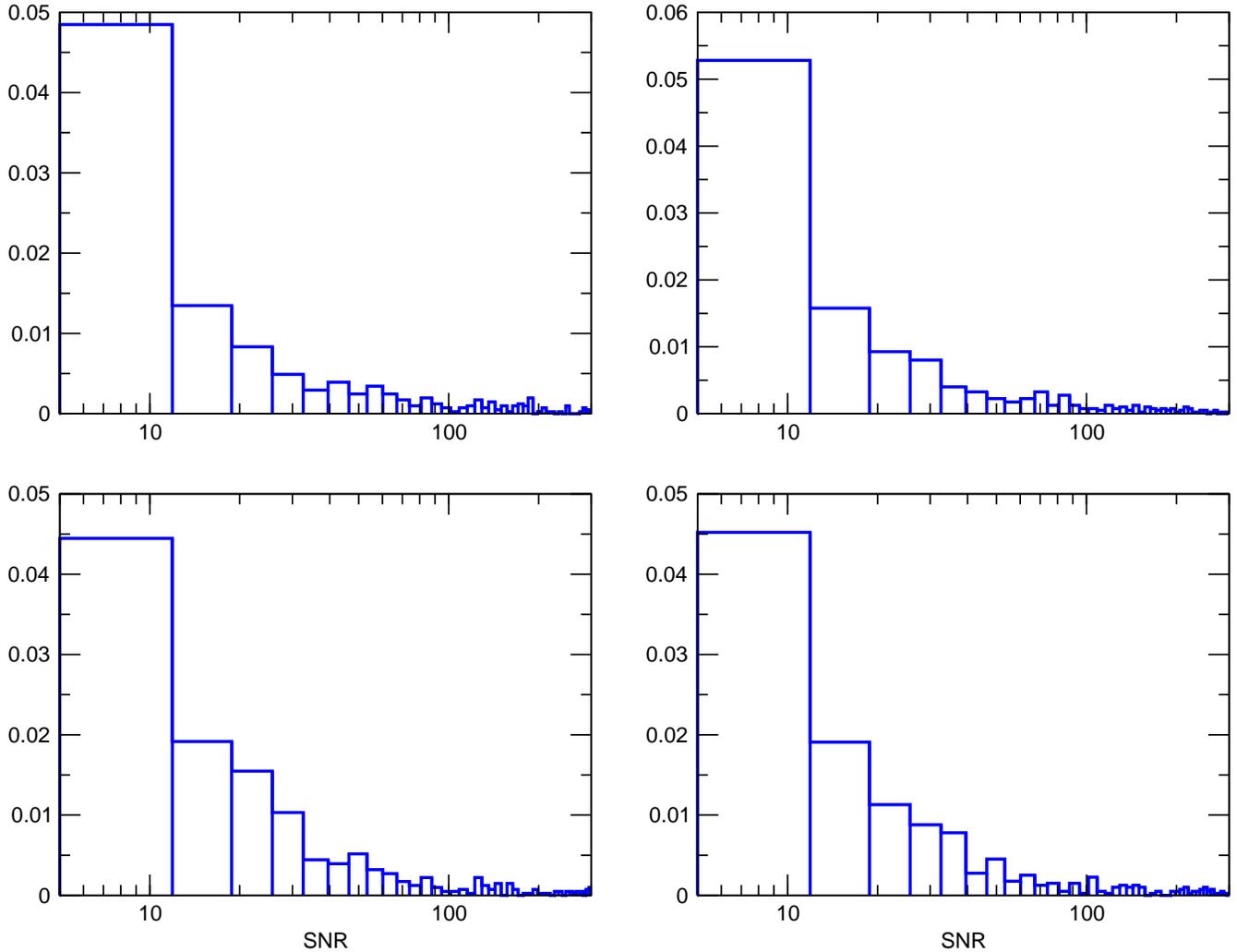}}
\caption{
The optimal SNRs for the inspiral of IMBH binary of Models~${\cal C, F, L}$ and
${\cal O}$ from the top to the bottom and from the left to the right. While it
is possible to have very strong sources with SNRs in the range of 300-400, the
vast majority of samples return SNRs of between 5 and 50 for all four models
}\label{fig:snrd}
\end{figure*}

\begin{figure*}
\resizebox{\hsize}{!}{\includegraphics[scale=1,clip]
{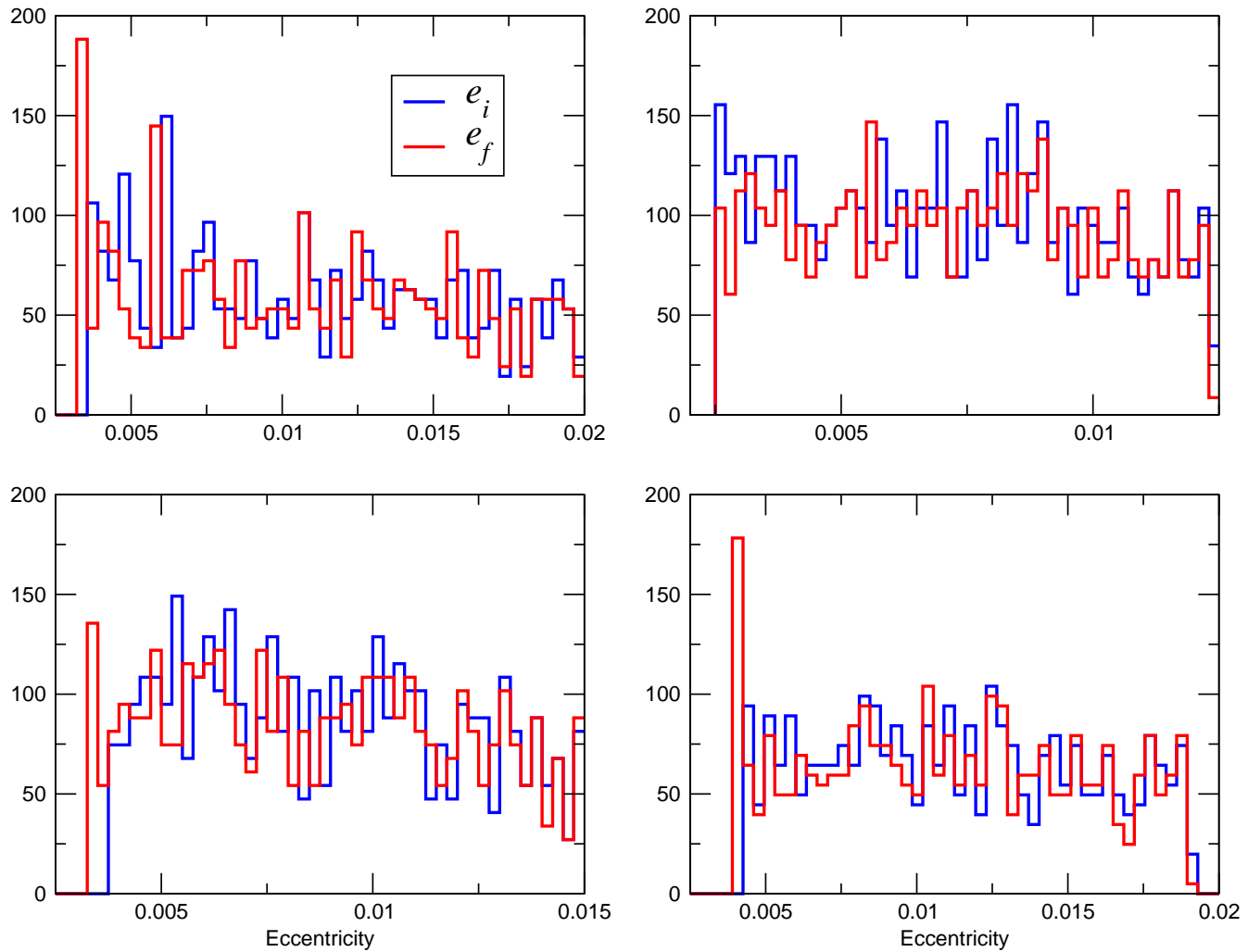}}
\caption{
The initial and final eccentricities for the inspiral of IMBH binary of
Models~${\cal C, F, L}$ and ${\cal O}$ from the top to the bottom and from the
left to the right, assuming a 3 year observation period for LISA.  As the
binary components are still quite widely separated, there is little
circularisation during the observation period, and the binaries thus retain a
measurable eccentricity. Note the different eccentricity scales in the different cells
}\label{fig:eccd}
\end{figure*}

\subsection{Results of the Monte Carlo}

In Figure~\ref{fig:snrd} we plot the recovered SNRs for the four models of IMBH
inspiral.  The models ${\cal C, F, L}$ and ${\cal O}$ are represented top to
bottom, and left to right.  We can see that while it is possible to have strong
sources, with SNRs $\ge 300$, the vast majority of samples returned more
modest, but detectable SNRs in the range of 5 to 50.  This means that IMBH
inspirals should be observable with the LISA detector.

Due to the lower mass ranges of these binaries, for systems placed at 100 Mpcs,
the sources start to become visible in the detector at GW frequencies of
$5\times10^{-4}$ Hz and higher. However, as they are still very widely
separated at this point, there is very little evolution in frequency or
eccentricity over a three year period.  In Figure~\ref{fig:eccd} we plot the
initial and final eccentricity distributions for the four models.  We can see
that there is very little circularisation of the binaries during the
observation period, which means that systems entering the observable LISA band
with eccentricities of $\sim0.02$ will reach the end of the three years with
almost the same eccentricity. As a consequence, these sources should retain a
measurable eccentricity throughout the LISA mission lifetime. We discuss about
the consequences of these results on lower-frequency Astrophysics and Data
Analysis in the next section.

\begin{table*}
\centering
\begin{tabular}{l|c|c|c|c}
  \hline
  $\Delta\vec{\lambda}$ & ${\cal C}$ & ${\cal F}$ &${\cal L}$
  & ${\cal O}$  \\
  \hline
 $\Delta M_c/M_c$  & $2.6982\times10^{-4}$ & $1.9155\times10^{-4}$ & $8.4848\times10^{-5}$ & $1.3321\times10^{-4}$ \\
 $\Delta \mu/\mu$  & $3.2393\times10^{-3}$ & $2.6694\times10^{-3}$ & $9.9361\times10^{-4}$ & $1.5502\times10^{-3}$ \\
 $\Delta D_L/D_L$  & $5.1611\times10^{-1}$ & $4.1823\times10^{-1}$ & $2.8644\times10^{-1}$ & $3.3223\times10^{-1}$ \\
 $\Delta\Omega$  & $2.1022\times10^{-3}$ & $1.7114\times10^{-3}$ & $9.4282\times10^{-4}$ & $1.1711\times10^{-3}$ \\
 $\Delta \cos\iota$  & $1.8954\times10^{-1}$ & $1.5333\times10^{-1}$ & $9.0724\times10^{-2}$ & $1.1194\times10^{-1}$ \\
 $\Delta e_0$  & $7.9722\times10^{-7}$ & $7.0795\times10^{-7}$ & $5.9972\times10^{-7}$ & $6.0454\times10^{-7}$ \\
 SNR  & $10.81$ & $12.68$ & $17.86$ & $15.31$ \\
  \hline
 \end{tabular}
\caption{
Median values of the parameter estimation errors and SNRs for the four models
${\cal C,F,L,O}$. Note that the units of $\Delta\Omega$ is steradians
}\label{tab:merrs}
\end{table*}

\section{Conclusions}
\label{sec.conclusions}

In this study we have carried out a dynamical study and a first step analysis
of the detection of IMBH binary systems in rotating clusters.  
For the case of a rotating King model without rotation, the results of the
presented survey verify previous outcomes by \citet{mak1993}, \citet{hem2002}
for massive black hole binary evolution in Plummer models facing the
development of the binding energy, the eccentricity and determined hardening
constants.  Analysing an extensive number of simulations, the main results 
from our study of the Dynamics of these systems can
be described: (1) The final eccentricity is strongly dependent on the initial
black hole velocities. (2) The eccentricity is dependent on the rotation
parameter of the model. (3) Determined hardening rates in the same range of
previous direct $N-$body simulations of comparable particle numbers. (4) Only
weak changes in the inclination and in the orientation of the angular momentum
vector direction have been observed, consistent with simulations by
\citet{mil2001}. (5) Counter rotation simulations yield noticeable different
results in eccentricity, in one case actually an extreme large value
$\bar{e}=0.997$. (6) Brownian motion of the centre of mass of the binary is
influenced by the rotation of the stellar system.
All simulations indicate that the orbital parameters eccentricity and
inclination develop to passably constant values in the non- or only weak bound
state, determined by initial conditions and the influence of dynamical
friction. 

In order to understand the impact of these sources in lower-frequency GW
Astrophysics, we have extended the direct $N-$body simulations with a
simplified semi-analytical model.  Whilst this approach is a ``kludge'' and can
only be envisaged as an approximation, the integration of the system down to
LISA's window is out of question because this would require months of CPU
calculation and, on the top of that, the numerical error would accumulate, so
that the results would not be as robust as what one can expect from
direct-summation schemes. 

We choose the systems yielding a larger eccentricity in the dynamical
simulations because these are the most appealing cases in the sense that their
detection will be very challenging. Also, information about the previous
dynamical story of the system is encoded in the radiation in the form of a
nonnegligible eccentricity.  

The results presented above show that LISA should have no problems in
identifying the existence of IMBH binaries. Such events are important for LISA
data analysis as they are a previously unconsidered source in terms of
dectability and parameter extraction. Our simulations also suggest that they
will spend their lifetime in the detector with a measurable eccentricity.  In
this work, we have looked at a particular case study where the masses and the
luminosity distance of the sources were fixed, and a Monte Carlo randomisation
carried out over the other response parameters.  We demonstrated that we will
be able to accurately measure the masses and sky resolution of such sources.
While the eccentricity is weak when the source becomes observable in the
detector, it should still be possible to carry out a precise measurement of the
initial eccentricity of the source.

For this we used the LFA response for the LISA detector.  This limited the
sources we investigated to a maximum GW frequency of 3 mHz to ensure that the
LFA was still valid.  As these are also quite high frequency sources, they have
a long generation time, which puts a time constraint of the size of the Monte
Carlo that can be carried out.  Finally, the waveforms used in this work
represent eccentric non-spinning binaries.

As well as detectability, the extraction of the system parameters is also of
great importance in GW Astrophysics. Using the FIM, we obtained the error
predictions for the important system parameters.  As the error predictions are
a function of the position of the source in the sky, plus the orientation of
the system with respect to the LISA constellation, the Monte Carlo simulations
produced error distributions with large tails.  Because of this fact, we have
decided to quote the median errors for the relevant parameters.   In
Table~\ref{tab:merrs} we present the median errors for the parameters $(M_c,
\mu, D_L, \Delta\Omega, \cos\iota, e_0)$.  We can see that for all models the
fractional errors in the estimation of the chirp-mass and reduced mass are of
the order of $10^{-4}$ and $10^{-3}$.  While there is not much frequency
evolution for these sources, the fact that they appear in the detector at
frequencies on the order of mHz means that we can obtain errors in the
luminosity distance of the order of $10^{-1}$.  We see a similar order of error
in the estimation of $\cos\iota$ which is in general a difficult quantity to
measure using electromagnetic information.

It is also quite remarkable to see that angular resolution of the IMBH
inspirals is very good, with median errors on the order of $10^{-3}$
steradians, corresponding to an error box on the sky of about 3 square degrees.
This level of accuracy would place an inspiralling IMBH firmly in the field of
view of a future detector such as the Large Synoptic Survey Telescope (LSST).
Finally, we can also see, again from the fact that these sources are emitting
GWs at frequencies on the order of mHz, the fractional errors in the estimation
of initial eccentricity is on the order of $10^{-7}$.

While we have shown that these IMBH binaries are detectable, there are a number
of ways in which the analysis can be improved in the future.  Firstly, a more
representative study would also have randomised the individual masses of the
binaries, as well as their luminosity distance.  This would allow us to give a
more concrete statement on detection and parameter estimation with the LISA
detector.  Secondly, we restricted the maximum GW frequency of the binary to 3
mHz to ensure a valid approximation to the LISA response.  In the future, we
could investigate higher frequency binaries by either using a Rigid Adiabatic
Approximation~\citep{rcp04} or full response to the LISA detector.  However, we
should point out that for the higher frequency binaries, the initial
eccentricity drops off rapidly, and these binaries may now be essentially
circular.  A more realistic study would also include the use of more realistic
waveforms which include spins and higher harmonics.  However, work on such
templates has yet to fully start in earnest.  Finally, it would also be
interesting to carry out a longer Monte Carlo, and assume different mission
lifetimes to see how detectability changes over observation time.

\section*{Acknowledgements}

PAS is thankful to Miguel Preto for his help with the rotinit subroutine to
create the initial conditions of the models.  The work of PAS has been
supported by the Deutsche Zentrum f{\"u}r Luft- und Raumfahrt at the Max-Planck
Institut f{\"u}r Gravitationsphysik (Albert Einstein-Institut) and by the
Ministerio de Educaci{\'o}n y Ciencia at the Institut de Ci{\`e}ncies de
l'Espai, IEEC/CSIC. PAS and EKP acknowledge the support of the Aspen Center for
Physics for the invitation in 2008. The simulations were partially performed at
the cluster {\sc Tuffstein} located at the AEI. PAS, EC and RS acknowledge
computing time on the GRACE cluster in Heidelberg (Grants I/80 041-043 of the
Volkswagen Foundation and 823.219-439/30 and /36 of the Ministry of Science,
Research and the Arts of Baden-W{\"u}rttemberg) The work was partially
supported German Science Foundation under SFB439 ``Galaxies in the Young
Universe'' and the Volkswagen Foundation.

\label{lastpage}

\end{document}